\documentclass[12pt,english]{article}

\usepackage[longnamesfirst]{natbib}
\usepackage{url}
\makeatletter
\def\url@leostyle{%
    \def\UrlFont{\sf}}{\def\UrlFont{\small\ttfamily}}
\makeatother
\usepackage{geometry}
\usepackage{fancyhdr}
\usepackage[font=small,labelfont=bf]{caption}

\usepackage[hidelinks,breaklinks]{hyperref}
\usepackage[hyphenbreaks]{breakurl}

\usepackage{authblk}

\usepackage{qcircuit}

\newcommand\xgate{%
  \Qcircuit @C=0.1em @R=.1em @! {
    & \gate{X} & \qw}}

\newcommand\ygate{%
  \Qcircuit @C=0.1em @R=.1em @! {
    & \gate{Y} & \qw}}

\newcommand\zgate{%
  \Qcircuit @C=0.1em @R=.1em @! {
    & \gate{Z} & \qw}}

\newcommand\rphasegate{%
  \Qcircuit @C=0.1em @R=.1em @! {
    & \gate{R} & \qw}}

\newcommand\sphasegate{%
  \Qcircuit @C=0.1em @R=.1em @! {
    & \gate{S} & \qw}}

\newcommand\hgate{%
  \Qcircuit @C=0.1em @R=.1em @! {
    & \gate{H} & \qw}}

\newcommand\cnotgate{%
  \Qcircuit @C=.1em @R=.1em @! {
    & \ctrl{1} & \qw \\
    & \targ & \qw}}

\usepackage{listings}
\usepackage{amsmath}

\usepackage{amssymb}

\usepackage{nicefrac}

\renewcommand*{\vec}[1]{\mathbf{#1}}

\usepackage{musixtex}

\newcommand*\fclef{%
  {\tiny \begin{music} \raisebox{1.2em}[2.5em][2.5em]{\bassclef} \end{music}}%
}
\newcommand*\gclef{%
  {\tiny \begin{music} \raisebox{0em}[3em][3em]{\trebleclef} \end{music}}%
}

\usepackage[T1]{fontenc}
\usepackage[charter]{mathdesign} 

\usepackage{graphicx}

\usepackage[english]{babel}

\numberwithin{equation}{section}

\usepackage{color}

\title{The Philosophy of Quantum Computing\thanks{%
    My work on this chapter benefited significantly from my interactions with students and other audience members during and after a series of lectures I gave at the University of Urbino's twenty-third international summer school in philosophy of physics, held online in June 2020 in the midst of the first wave of the COVID-19 pandemic, as well as a further lecture I gave for Michel Janssen's ``The Age of Entanglement'' honours seminar at the University of Minnesota in December 2020 as the second wave of the pandemic began in earnest. Thanks to Ari Duwell, Eduardo Miranda, Philip Papayannopoulos, and Lev Vaidman for comments on a previous draft of this chapter. I am also grateful for informal discussions, over the years, with Guido Bacciagaluppi, Jim Baggot, Michel Janssen, Christoph Lehner, Lev Vaidman, and David Wallace; my presentation of the Everett interpretation in Section \ref{s:many-worlds}, in particular, is significantly informed by what I take myself to have learned from these discussions, though I hold only myself responsible for any mistakes or misunderstandings in my presentation of the Everettian view. Section \ref{s:fundamentals}, on ``Fundamental concepts'' is heavily informed by my recent work on related topics with Stephan Hartmann, Michael Janas, Michel Janssen, and Markus M\"uller, as well as by informal correspondence with Jeffrey Bub and (the late) Bill Demopoulos; though here again I take sole responsibility for any mistakes. Finally, I gratefully acknowledge the generous financial support of the \emph{Alexander von Humboldt Stiftung}.}}
\author[a]{Michael E. Cuffaro}
\affil[a]{{\small Munich Center for Mathematical Philosophy, Ludwig-Maximilians-Universit\"at M\"unchen}}
\date{}

\begin{document}

\maketitle

\thispagestyle{empty}

\begin{center}
\emph{To appear in the book Quantum Computing in the Arts and Humanities: An Introduction to Core Concepts, Theory and Applications. E. R. Miranda (Ed.). Cham: Springer Nature, 202x. Pre-publication draft v.16 Mar 2021.}

\mbox{ } \\
\end{center}

\section{Introduction}
\label{s:intro}

From the philosopher's perspective, the interest in quantum computation stems primarily from the way that it combines fundamental concepts from two distinct sciences: physics (especially quantum mechanics) and computer science, each long a subject of philosophical speculation and analysis in its own right. Quantum computing combines both of these more traditional areas of inquiry into one wholly new (if not quite independent) science. There are philosophical questions that arise from this merger, and philosophical lessons to be learned. Over the course of this chapter we will be discussing what I take to be some of the most important.\footnote{Space does not permit me to exhaustively survey all of the philosophical issues brought up by quantum computing. The interested reader can find a summary of other important issues in \citet[]{hagar2019}.}

We begin, in Section \ref{s:fundamentals}, by introducing the fundamental concepts from physics and computation that will be essential for framing the further philosophical discussion that will follow. Section \ref{s:classical-states} and Section \ref{s:classical-computers} introduce concepts from classical mechanics and the classical theory of computation, respectively. In Section \ref{s:classical-states} we discuss the concept of the state of a physical system as it is given in classical mechanics. We emphasise in particular the way that we are invited, in classical mechanics, to think of the state of a physical system as a compact description of the properties possessed by it. These properties determine, in advance, the answers to the experimental questions that we can pose of a system. And considering these questions and their answers leads one to an alternative representation of a system's state that is useful for representing computation, a subject we then take up in more detail in Section \ref{s:classical-computers}.

In Section \ref{s:classical-computers} we introduce the concept of a model of computation (or computational architecture), the concept of the cost of carrying out a computation under a given computational model, and explain what it means to solve a problem just as easily under one computational model as under another. After discussing some of the more important computational complexity classes, we then introduce two theses that can be used to ground the model-independence of computational cost: the so-called universality of Turing efficiency thesis (sometimes also referred to as the `strong', or `physical', or `extended' Church-Turing thesis), and the invariance thesis. These theses are both called into question by the existence of quantum computers. And since some have taken the absolute model-independence guaranteed by them to be foundational for the science of computational complexity theory, the question arises of what to make of those foundations in light of the existence of quantum computation. We will discuss this question, in an abstract way, in Section \ref{s:pccp0}, where I will argue that although the universality thesis must indeed be given up, this is not a great loss. The invariance thesis, by contrast, remains, but in a different form, transformed from a metaphysical to a methodological principle.

In Section \ref{s:quantum-states-and-operations} we turn to the physics behind quantum computation, and begin by introducing some of the more important concepts of quantum theory. Most importantly we introduce the concept of a quantum state. We emphasise that the way that a quantum state determines what the answers to experimental questions will be is fundamentally different than the way that they are determined by a classical state. We then turn to quantum computation proper in Section \ref{s:many-worlds}, where we review the basic concepts of quantum computing and consider what we can say regarding the physical explanation of the power of quantum computers. The many-worlds explanation of quantum computing---the idea that quantum computers outperform classical computers by running their computations in exponentially many physical universes---is then introduced. We note two major problems that arise for this interpretation. The first arises from the so-called preferred basis problem. This problem is a challenge (that is arguably surmountable) for the many-worlds view in the more general context of quantum mechanics. But we will see that it is especially problematic in the context of quantum computers. The second major problem arises from the fact that there are many different models of quantum computation, but the many-worlds explanation of quantum computing only seems motivated by one of them.

In Section \ref{s:entanglement} we consider the role of quantum entanglement in enabling quantum computers to outperform classical computers. We summarise an unsuccessful argument to the conclusion that quantum entanglement is insufficient to enable this quantum `speedup' in Section \ref{s:gk-theorem}, noting in Section \ref{s:computer-simulations} that reflecting on what it means to provide a classical computer simulation of a quantum phenomenon should convince us to reach the opposite conclusion. We continue the discussion of classically simulating quantum computers in Section \ref{s:pccp1}, and reflect on general aspects of the computational perspective that the study of quantum computing provides on physical theory. We note that reflecting on quantum computing emphasises that there are important differences in the methodological presuppositions that lie at the basis of physics and computer science, respectively, and that conflating these can lead to confusion. We also note the emphasis that studying quantum computation places on the fact that quantum mechanics and classical mechanics are each alternative universal languages for describing physical systems, and that the difference between quantum mechanics and classical mechanics lies, fundamentally, in the differences in the expressive power of these languages.\footnote{\label{fn:framework}By `quantum mechanics' I mean the fundamental theoretical framework shared in common by every specific quantum-mechanical theory (quantum field theories, for instance) of a particular class of systems; see \citet[pp. 110--111]{aaronson2013}, \citet[][ch. 1 and \S 6.3]{3m2020}, \citet[][p. 2]{nielsenChuang2000}, and \citet[]{wallace2019}.} We reflect on this in Section \ref{s:conclusion}.

\section{Fundamental concepts}
\label{s:fundamentals}

\subsection{Classical states}
\label{s:classical-states}

Classical mechanics (in its various forms\footnote{These include Newtonian, Lagrangian, Hamiltonian, relativistic, and classical statistical mechanics. For a recent comparison and philosophical discussion of Lagrangian and Hamiltonian mechanics, see \citet{curiel2014}.}) was, prior to quantum mechanics, our fundamental theoretical framework for describing the dynamics of physical systems (i.e., how physical systems change over time). What is meant by a physical system here is just one of the concrete objects that a particular physical theory describes. All such systems are described as evolving through time in accordance with the dynamical constraints that a theoretical framework applies universally to every physical system. In classical mechanics (and the same is true, as we will see later, in quantum mechanics) a physical system can be anything from a single particle to (in principle) the entire physical universe. Mathematically, though, what the dynamical laws of a theory actually govern are the relations between the possible state descriptions of systems. By the state description (or state specification) of a system is meant a description of the particular physical situation that it happens to be in at a given moment in time. For instance, at a given moment in time, a classical-mechanical system will have a particular kinetic energy, it will be accelerating (or not) in a particular direction, and so on.

It turns out that we do not have to explicitly specify each and every one of a system's dynamical properties to exactly and exhaustively specify the system's dynamics. In classical mechanics, given the dynamical laws, it is enough to specify values for a system's position and momentum. The value of any other dynamical property can then be calculated by relating these with each other and with the values of non-dynamical properties of the system such as, for instance, its charge or its mass.

Figure \ref{f:compact-state} illustrates the two different strategies described in the previous paragraph for specifying the state of a physical system, for instance a system composed of a particle attached to the end of a spring constrained to move in one direction \citep[see][p. 73]{hughes1989}. On the left we explicitly specify values ($v_1$, $v_2$, $\dots$) for each and every one of the physical system's dynamical parameters. On the right the values of momentum, $p$, and position, $q$, are specified, and all other dynamical quantities are calculated on their basis. In particular the total energy, $H$, of the system is defined as the sum of the kinetic energy, $T$, of the particle, and the potential energy, $V$, stored in the spring, which are in turn defined in terms of $p$ and $q$, respectively. Note that $m$, the system's mass, is a \emph{non}-dynamical parameter that does not change over the history of the system, and $k$ is the force (the first derivative of momentum) per unit distance required for the spring to be displaced from its equilibrium point (i.e., the point at which it is neither stretched nor compressed). Other forces (e.g., gravitational forces) are neglected but in principle can be included \citep[for further discussion, see][\S 3.2.5]{hughes1989}.

\begin{figure}
  \begin{center}
    \begin{tabular}{ l l | l }
      $P_1 ~=~ v_1$ & & $\quad p ~=~ v_p$ \\
      $P_2 ~=~ v_2$ & & $\quad q ~=~ v_q$ \\
      $P_3 ~=~ v_3$ & & $\quad T ~=~ \nicefrac{p^2}{2m}$ \\
      $P_4 ~=~ v_4$ & & $\quad V ~=~ \nicefrac{kq^2}{2}$ \\
      $P_5 ~=~ v_5$ & & $\quad H ~=~ V + T$ \\
      $\dots$       & & $\quad \dots$
    \end{tabular}
  \end{center}

  \caption{Two different strategies for specifying the state of a physical system, for instance a system composed of a particle attached to the end of a spring constrained to move in one direction \citep[see][p. 73]{hughes1989}. On the left we explicitly specify values ($v_1$, $v_2$, $\dots$) for each and every one of the physical system's dynamical properties. On the right the values of momentum, $p$, and position, $q$, are specified, and all other dynamical quantities are calculated on their basis.}
\label{f:compact-state}
\end{figure}

The upshot is that specifying the position and momentum of a system provides us with enough information, in classical mechanics, to completely characterise all of the other dynamical properties of that system. Accordingly, the dynamical state, $\omega_t$, of a system at some particular time $t$ is given in classical mechanics by:
\begin{align}
  \label{e:omega}
  \omega_t = (\vec{q}_t, \vec{p}_t),
\end{align}
where $\vec{q}_t$ and $\vec{p}_t$ are vectors (in three dimensions) representing the system's position and momentum, respectively, at $t$. Further, we can infer from the state of a system at $t$, and the classical-mechanical laws of motion, exactly what the state of the system will be in the next moment and at every other time both in the system's past and in its future \citep[\S 2.6]{hughes1989}.

Classical-mechanical states have another feature. Imagine all of the possible (experimental) yes-or-no questions one might want to ask about the dynamical properties of a particular system at a particular time, questions like: \emph{Is the value of the dynamical property $A$ within the range $\Delta$?} Completely specifying a system's dynamical state, i.e., specifying its momentum, $\vec{p}$, and its position, $\vec{q}$, yields a simultaneous answer to all such questions irrespective of whether any question has actually been asked. Indeed, this is what actually justifies our having considered the values of $\vec{p}$, $\vec{q}$, and the other quantities derived from them to be (observable) \emph{properties} of the system in the first place, properties possessed by the system whether or not, and however, we enquire concerning them \citep[\S 6.3]{3m2020}. The same goes for a system composed of many parts; for after all, any observable property $A_1$ of a subsystem $S_1$ of some larger system $S$ is also an observable property of $S$. Thus, given the state specification for a particular system we can construct a sort of `truth table' which could be used, in principle, to answer any conceivable yes-or-no question concerning the system and its subsystems (see Figure \ref{f:truth-table}).

\begin{figure}
\footnotesize
\begin{align*}
  \begin{array}{l | l | l | l || l | l | l }
    \vec{p}_1 & \vec{q}_1 & \vec{p}_2 & \vec{q}_2 & A & B & \dots \\[2pt] \hline
    v^1_{p_1}   & v^1_{q_1}   & v^1_{p_2}   & v^1_{q_2}  & N & N & \\[2pt] \hline
    v^2_{p_1}   & v^2_{q_1}   & v^2_{p_2}   & v^2_{q_2}  & N & Y & \\[2pt] \hline
    v^3_{p_1}   & v^3_{q_1}   & v^3_{p_2}   & v^3_{q_2}  & N & Y & \\[2pt] \hline
    v^4_{p_1}   & v^4_{q_1}   & v^4_{p_2}   & v^4_{q_2}  & Y & N & \\[2pt] \hline
    v^5_{p_1}   & v^5_{q_1}   & v^5_{p_2}   & v^5_{q_2}  & N & Y & \\[2pt] \hline
    v^6_{p_1}   & v^6_{q_1}   & v^6_{p_2}   & v^6_{q_2}  & Y & N & \\[2pt] \hline
    v^7_{p_1}   & v^7_{q_1}   & v^7_{p_2}   & v^7_{q_2}  & Y & N & \\[2pt] \hline
    v^8_{p_1}   & v^8_{q_1}   & v^8_{p_2}   & v^8_{q_2}  & N & N & \\
  \end{array} \\
  \begin{array}{c c c c c c}
    \dots & \mbox{ } & \mbox{ } & \mbox{ } & \mbox{ } & \mbox{ }
  \end{array}
\end{align*}
\caption{A `truth table' for a classical system composed of two subsystems whose states, respectively, are $(\vec{p}_1, \vec{q}_1)$ and $(\vec{p}_2, \vec{q}_2)$. Various combinations of values for the state parameters $\vec{p}_i$ and $\vec{q}_i$ are given on the left of the double-vertical line, with the superscript $j$ in $v^j$ indicating the $j^{th}$ combination. Relative to a given combination of values, the answers to experimental questions concerning the values of derived observable quantities $A$, $B$, $\dots$ on the right of the double-vertical line can be determined.}
\label{f:truth-table}
\end{figure}

\begin{figure}
  \begin{center}
    \includegraphics[scale=0.5]{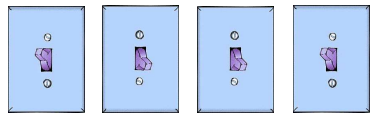}
  \end{center}
  \caption{A physical system consisting of four light switches. For each switch, we can ask: ``Is switch $S$ turned on?'' Using 0 to represent yes and 1 to represent no, the state of the overall system relative to this question can be represented using the bit-string 0110 (which, in base-ten, is the number 6). This is just one of the $2^4 = 16$ possible states that a system like this one can be in. More generally, for a system made up of $n$ 2-dimensional subsystems, the number of possible states for the system as a whole is $2^n$.}
  \label{f:switches}
\end{figure}

\begin{figure}
  \begin{center}
    \qquad \includegraphics[scale=.4]{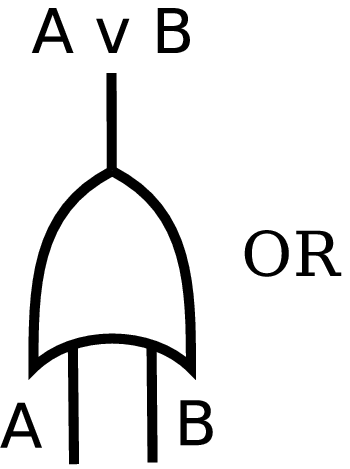} \qquad \qquad \includegraphics[scale=.4]{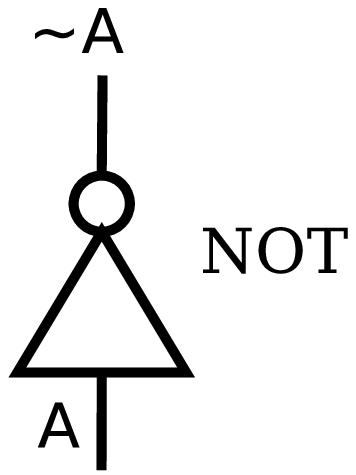} \qquad \includegraphics[scale=.4]{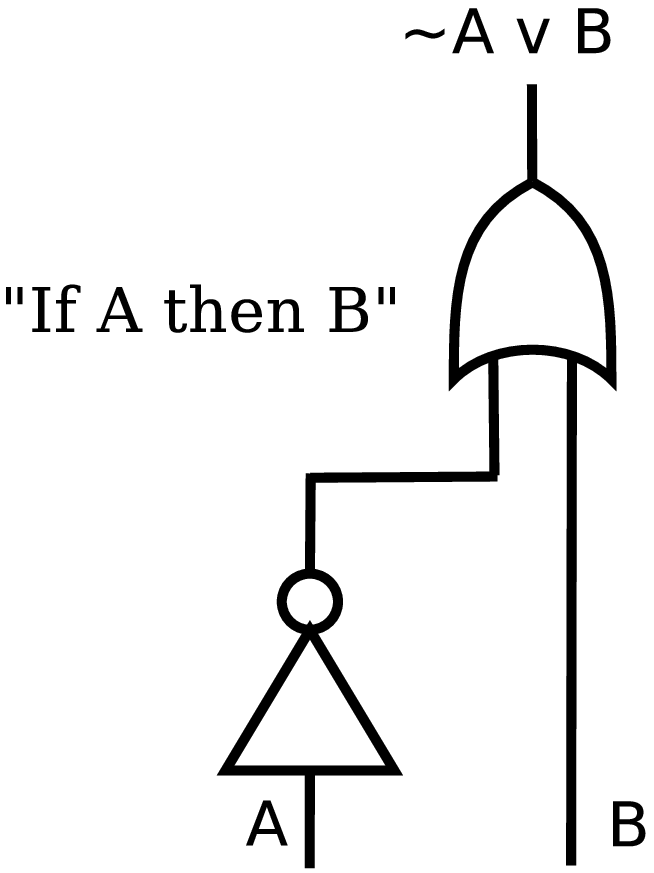}
  \end{center}
  \caption{Some of the logico-mathematical operations, or `logic gates', that can be used to manipulate bits. OR and NOT together constitute a universal set of gates, in the sense that any more complicated logico-mathematical operation, such as the `if-then' gate at the right, can be constructed using combinations of OR and NOT. There are other universal sets besides \{OR, NOT\}, for instance: \{AND, NOT\}, \{NOR\}, and \{NAND\}, where NAND is the `not-and' gate and NOR is the `not-or' gate.}
  \label{f:classical-gates}
\end{figure}

In Figure \ref{f:truth-table}, I wrote $Y$ and $N$ (for ``yes'' and ``no''), but I could have equally well used $T$ (for true) and $F$ (for false), or alternately the binary digits 0 and 1. Using binary digits, i.e., `bits', is convenient because they are useful for representing numbers, which can be manipulated abstractly using logico-mathematical operations.\footnote{I say `logico-mathematical' because logical operations on bits can be thought of as modulo-2 arithmetical operations \citep[see][]{boole1847}.} In particular, if we build physical systems whose states can reliably be used to represent binary numbers (in the sense that yes-or-no questions concerning their observable properties $A$, $B$, $\dots$ can be laid out as in Figure \ref{f:truth-table}, replacing 0 for $Y$ and 1 for $N$), and reliably evolve them in ways that mirror a small basic set of logico-mathematical operations, then we can (by combining these operations) use physical systems to carry out computations that can in principle be arbitrarily complex (see Figures \ref{f:switches}--\ref{f:classical-gates}).\footnote{For more general accounts of how physical systems can be used to represent computations, see \citet[]{fletcher2018, horsman2018, maroney2018}.}

\subsection{Classical computers}
\label{s:classical-computers}

When the word `computer' is used in popular discourse today, generally what is meant is a physical system like the one depicted in Figure \ref{f:switches}---\emph{not}, of course, literally a collection of light switches, but a physical system composed of subsystems that, like the switches, can reliably be interpreted as being either on or off, and that can be organised into registers, random access memory, hard drives, and other peripheral systems such as keyboards, `mice', touchpads, and so on, that can be manipulated to reliably realise various logical operations like the ones depicted in Figure \ref{f:classical-gates}. The laptop I am writing this chapter with is an example of such a device. We turn such devices on when we need them, and turn them off or `put them to sleep' when we do not. But the word `computer' did not always signify a kind of machine. Prior to the 1930s, `computation' generally meant an activity carried out by a human being. Even after Alan Turing's\nocite{turing1937} introduction of what later came to be known as the `Turing machine' in 1936---an abstract device that is itself modelled on human computation---`computers' were generally understood to be human beings until well into the 1960s.\footnote{The role of human computers in the United States of America's space program, for instance, has been documented in \citet[]{shetterly2016}.}

\begin{figure}[t]
\begin{center}
  \includegraphics[scale=0.5]{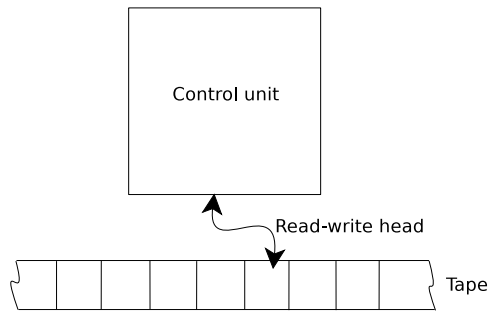}
  \caption{A version of what is now called a `Turing machine', with a single bi-directional tape, a control unit encoding a particular set of instructions or program, and a read-write head. Such a machine is an idealised representation of the components involved in human computation. In a \emph{universal} Turing machine, the computer's program is read from the tape at the beginning of the computation. For a good introduction to the Turing machine and other related topics, see \citet[]{martin1997}.}
\label{f:dtm}
\end{center}
\end{figure}

\sloppypar{As for the Turing machine: this, as mentioned, is also not a concrete physical device (although it could, in principle, be instantiated by one) but an abstract mathematical model. Turing's primary reason for introducing it had been to address the so-called \emph{Entscheidungsproblem} (which is the German word for `decision problem'), an important but unsettled question that had arisen in the foundations of number theory. The \emph{Entscheidungsproblem} concerned the existence of an effective procedure for deciding whether an arbitrarily given expression in first-order logic can be proven from the axioms of first-order logic. The answer, as Turing and, independently, Alonzo \citeauthor[]{church1936} were able to demonstrate, is that there is no such procedure. Turing's own proof relied on a penetrating philosophical analysis of the notion of effective (human) computation, and a corresponding argument that one could design an automatic machine (now called a \emph{Turing machine}; see Figure \ref{f:dtm}) to carry out the essential activities of a human computer.

\citet[pp. 249--51]{turing1937} argued that, for a computer to carry out a computation, it is essential that she have access to a notebook from which she can read and onto which she can write various symbols in the course of her work. These symbols need to be distinguishable by her from one another, on the basis of which Turing argued that the alphabet from which she chooses her symbols must be a finite alphabet. At any given moment during a computation, a computer may find herself in one of a finite number (again, because she must be able to distinguish them) of states of mind relevant to the task at hand which summarise her memory of the actions she has performed up until that point along with her awareness of what she must now do (pp. 253--4). The actions that are available to her are characterised by a finite set of elementary operations, such as `read the next symbol' from the notebook, `write symbol $a$' to the notebook, and so on. In an automatic machine, a \emph{control unit} is constructed to encode the machine's `state of mind' (i.e., a logical representation of its current state, its current input, and its transition function), which in general changes after every operation of the \emph{read-write head}. The latter moves back and forth along a \emph{one-dimensional tape} (the machine's `notebook'), from which it reads and onto which it writes various symbols from a finite alphabet, in conformity with a particular finite set of exactly specifiable rules. Turing was able to show that no automatic machine of this kind can be used to solve the \emph{Entscheidungsproblem}.

Through the work of Turing, Church, and others, the modern science of computation, and in particular \emph{computability theory}---the science of which problems can and which cannot be solved by a computer, i.e., by any thing or any person that can carry out (but is restricted to) the essential activities associated with human computation---was born.}\footnote{For more on the \emph{Entscheidungsproblem} and the early history of computer science, see \citet[]{copeland2020, davis2000, dawson2007, lupacchini2018}.}

Irrespective of whether it is a machine or a human being that is doing the computing, the question of how much it actually costs to compute a solution to a computable problem is also a very important one. In particular we generally would like to know how much \emph{time} and how much \emph{space} are required (though time is usually regarded as the more important measure). The question of the cost of a given computation had become especially important as more and more problems came to be carried out by machines in the 1950s and 1960s. Even the earliest digital computers of the 1930s and 1940s performed their computations far faster than human beings could possibly perform them, and with improvements in design and in technology, machines could be made to run ever faster. Through the need to use these machines' resources efficiently, a sub-field of computer science began to develop whose concern was with how efficiently a given problem can be solved and on which types of machine.

In the context of the Turing machine model, time and space are quantified in terms of the number of computational `steps' (i.e., the number of basic operations) and the number of squares of tape, respectively, that are needed to carry out a given computation. Although it is in principle possible (assuming we have enough tape) to instantiate a Turing machine physically,\footnote{A Turing machine's tape does not need to be actually infinite in length. What is required is only that the tape be \emph{indefinitely} long, so that, for a given (finite) computation, the machine can be supplied with enough tape to carry it out. To put it another way: What constitutes `enough tape' to carry out a computation is not part of the general definition of a Turing machine. It is, rather, assumed, for the purposes of that definition, that enough tape to carry out a given computation can be supplied. That said, certain variations on the Turing machine model restrict the ways in which tape can be read by the control unit in various ways. For instance some variants employ separate tape(s) for the machine to write ``rough work'' on in addition to an output tape, some variants only allow the read-write head to move in one direction along the tape, and so on.} generally it makes more sense, from a practical perspective, to design a mechanical computer in accordance with some other model of computation. Particularly important is the von Neumann architecture, a type of stored-program computer,\footnote{Although the von Neumann architecture, or `von Neumann machine' is only one of a number of various types of stored-program computer, the terms have today (inappropriately, from a historical point of view) come to be understood synonymously \citep[]{copeland2017}.} that involves a central processing unit (CPU), an instruction register and program counter, internal memory used to store data (including the program being run), external memory used to store less frequently needed data, and finally an input and an output mechanism. A von Neumann machine can in principle compute all of the same problems as a universal Turing machine (see Figure \ref{f:dtm}), but a time step for a von Neumann machine is not concretely interpreted in the same way as it is for a Turing machine, just as `space' for a von Neumann machine is not measured in squares of tape.

There are many other universal models of computation besides von Neumann machines and Turing machines. Nevertheless, at least when it comes to (reasonable) classical computational models, it is possible to talk about the computational cost of solving a particular problem (like, for instance, the problem of finding the prime factors of a given integer), in a model-independent way; i.e., without having to specify which machine model that cost is defined with respect to. There are two things involved in explicating what it means for the computational cost of solving a given problem to be model-independent in this sense. The first, mentioned just now in passing, is the idea of a reasonable machine architecture.

Roughly, a \emph{reasonable} (universal) machine architecture or model is one that it would be possible to physically instantiate. Turing machines like the one described in Figure \ref{f:dtm} and von Neumann machines are examples of reasonable machine models. A computational model that employs an unbounded amount of parallel processing, by contrast, is an example of an unreasonable model, or what Peter \citet[p. 5]{boas1990} calls a model of the `second machine class'. There are legitimate theoretical reasons for studying the complexity-theoretic properties of unreasonable models. But no finite physical system could possibly instantiate an unreasonable architecture. There are thus good reasons to want to distinguish the reasonable from the unreasonable ones \citep[see also][]{dean2016b}.

The second thing we need to do, in order explicate what it means for the computational cost of solving a given problem to be model-independent, is to make sense of what it means to solve a given problem \emph{just as easily} with one computational architecture as with some other one. But before we can make sense of solving something just as easily we need to first try and make sense of what it means to solve something \emph{easily} under one computational model. Consider, to begin with, a human computer working alone. For a human computer, the difference between a problem that could in principle require up to a thousand steps to solve, and one that could in principle require up to a million steps, is very great indeed. With some justification this person might call the former problem easy to solve, and the latter problem hard. Any criterion based on a fixed number of computational steps, however, would not be very useful in the context of machines, at any rate certainly not any criterion based on numbers this low. Even by the 1950s a mechanical computer could carry out a computation far faster than a human being. As of 2020, a typical laptop computer can perform hundreds of thousands of Millions of Instructions per Second (MIPS) \citep[]{wikiMips}, and computers are becoming more and more performant every year. In the context of machine computation, it is far more useful to ask how a given solution to a computational problem \emph{scales}; i.e., how the number of resources required to carry out a given solution to that problem grows with the size of the input to the problem. Take, for instance, the problem to sort a given set of numbers. All else equal, the more numbers there are to sort, the longer it will take. How much longer? The fastest sorting algorithms (one of which is {\ttfamily MergeSort}) will take on the order of $n\log n$ steps to sort $n$ numbers in the worst case \citep[]{mehlhorn2008}.

In modern computational complexity theory, we say that a problem is \emph{easy} to solve, or alternately that it is \emph{tractable}, if there exists an efficient algorithm to solve it. If no efficient algorithm exists, then we say that the problem is \emph{hard}, or \emph{intractable}. This definition, of course, merely trades in one informal notion for another, so let us state more precisely what we mean by `efficient'. By an \emph{efficient algorithm} for solving a given problem, we mean an algorithm that will solve it in \emph{polynomial time}; i.e., in a number of time steps that is no more than a polynomial function of the size of its input. Stated in mathematical terms, this means that for input size $n$, such an algorithm will take on the order of $n^k$ steps, where $k$ is some constant.\footnote{\label{fn:big-oh}`On the order of' is a technical term, usually symbolised in ``big-oh notation'' as $O(T(n))$. An algorithm is $O(T(n))$ for some function $T(n)$ if for every sufficiently large $n$, its actual running time $t(n) \leq c \cdot T(n)$ for some constant $c$. For instance, an algorithm that never takes more than $5n^3$ steps is $O(n^3)$.} As for an algorithm that takes more than a polynomial number of steps, we consider it to be \emph{inefficient}, even if it might be the most efficient algorithm known for a particular problem. An example of an inefficient algorithm is one that requires \emph{exponential time}; i.e., on the order of $k^n$ steps for input size $n$ and some constant $k$.\footnote{A famous example of a problem for which only exponential-time solutions are known is the Travelling Salesman Problem \citep[]{cook2012}.} Even harder problems take \emph{factorial time}, i.e., on the order of $n!$ steps, and so on.

The idea of using polynomial time as our criterion for distinguishing between efficient and inefficient algorithms---what we will henceforth call the \emph{polynomial principle}---was introduced independently by Alan \citeauthor[]{cobham1965} and Jack \citeauthor[]{edmonds1965} in 1965.\footnote{In some literature it is referred to as the \emph{Cobham-Edmonds thesis}. Kurt G\"odel anticipated the principle, to some extent, in a private letter he wrote to John von Neumann in \citeyear{godel1956}. For further discussion, see \citet[]{cuffaro2018b}.} There are a number of good reasons for adopting the polynomial principle. One reason is simply that the set of problems picked out by it has so far tended to correspond with those that we would (so to speak) pre-theoretically regard as efficiently solvable \citep[\S 11.6]{cuffaro2018b}.\footnote{This correspondence is not perfect, but the usefulness of the polynomial principle is such that we appeal to it despite this \citep[\S 11.6]{cuffaro2018b}.} Perhaps the most useful feature of the polynomial principle stems from the fact that polynomial functions compose (see Figure \ref{f:polynomial}). That is, given an algorithm that runs in polynomial time, if we add to it any number of calls to polynomial-time subroutines, the total running time of the algorithm will still be polynomial \citep[p. 27]{arora2009}.\footnote{It is easy to see this: Consider a program that consists of $n^k$ calls of a subroutine that takes $n^l$ steps, where $n$ is the number of bits used to represent the input, and $k$ and $l$ are finite constants. The total number of steps needed to carry out this program will be $n^{k+l}$. If $k$ and $l$ are finite constants then so is $k+l$. In other words, $n^{k+l}$ is still a polynomial.}

\begin{figure}
  \begin{lstlisting}[]
    // A polynomial-time algorithm:         // Another polynomial-time algorithm:
    myProgram1(arg1, arg2) {                 myProgram2(arg1, arg2) {
      statement1;                             statement1;
      statement2;                             statement2;
      ...                                     ...
    }                                         myPolynomialTimeSubroutine(arg1);
                                              ...
                                              myPolynomialTimeSubroutine(arg2);
                                            }
  \end{lstlisting}

  \caption{Inserting any number of calls to polynomial-time subroutines within the body of a polynomial-time algorithm results in a polynomial-time algorithm.}
  \label{f:polynomial}
\end{figure}

Now that we have a handle on what it means for a problem to be solvable easily, we can state what it means for a problem to be solvable just as easily on one type of computational architecture as on another. We say that a problem is solvable \emph{just as easily} on machine model $\mathfrak{M}_1$ as on machine model $\mathfrak{M}_2$ if and only if there is an algorithm for solving the problem on $\mathfrak{M}_1$ that requires no more than a polynomial number of extra time steps as compared to $\mathfrak{M}_2$.\footnote{Note that it makes sense to talk about solving a given problem just as easily on $\mathfrak{M}_1$ as on $\mathfrak{M}_2$ even when the problem under consideration is actually intractable for both. For instance if some problem requires $2^n$ steps to solve on $\mathfrak{M}_1$ and $2^n + n^3$ steps to solve on $\mathfrak{M}_2$ then it is no harder, from the point of view of the polynomial principle, to solve it on $\mathfrak{M}_2$ than on $\mathfrak{M}_1$.} Another way of putting this is that anything that $\mathfrak{M}_2$ can do is \emph{efficiently simulable} by $\mathfrak{M}_1$.

Any problem that is solvable in polynomial time by a Turing machine is said to belong to the class \textbf{PTIME} (for `polynomial time'), though it is usually referred to simply as \textbf{P}. \textbf{P} is, of course, not the only complexity class. If a given problem is such that, for a given solution to it, there is a Turing machine that will \emph{verify} that the solution is indeed a solution, then the problem is said to belong to the class \textbf{NP}, for `nondeterministic polynomial time'. The reason for the name is that, equivalently, this class of problems can be defined as those that can be solved in polynomial time on a nondeterministic Turing machine. A \emph{nondeterministic Turing machine}, or \emph{choice machine}, is such that at a given step of a given computation, an imagined external operator of the machine can choose to have it transition in one way rather than another. This is unlike a standard (deterministic) Turing machine for which every computational step is completely determined in advance given a particular input. Choice machines are interesting from a theoretical point of view, but given that they involve an external operator they are not really automatic machines in the sense that is relevant to our discussion here.

A variant of the choice machine that \emph{is} automatic is a \emph{probabilistic Turing machine}. Such a machine's choices are not made by an external operator but by the machine itself, which we can imagine as equipped with the equivalent of a fair coin that it can flip a number of times to decide whether to transition one way or another whenever such a choice is open to it.\footnote{For more on probabilistic and nondeterministic Turing machines and how they compare to their deterministic counterparts, see \citet[\S 11.3]{cuffaro2018b}.} The class of problems solvable in polynomial time on a probabilistic Turing machine is called \textbf{BPP} (for `bounded error probabilistic polynomial'). Note that what it means to be able to solve a given problem is not the same for a probabilistic Turing machine as it is for a deterministic Turing machine. In particular a given `solution' output by a probabilistic Turing machine is allowed to be wrong. We only demand that it be right with high enough probability so that, if we re-run the machine for the given input on the order of a polynomial number of further times, our confidence in the majority answer will approach certainty.

There are very many more complexity classes besides these ones \citep[see][]{aaronson2012}. But we will stop here as we now have the basic ingredients with which to state a couple of related theses that can be used to explicate what it means for the cost of solving a given problem to be model-independent. As we will see a little later, there are good reasons to prefer the second thesis, but for now we simply introduce both. First, the \emph{universality of Turing efficiency thesis} (sometimes also called the `strong', `physical', or `extended' Church-Turing thesis\footnote{See \citet[ch. 6]{timpson2013} for discussion of a different, more general thesis, that is only indirectly relevant to computational complexity theory. For a discussion of how these theses relate, see \citet[\S 11.4]{cuffaro2018b}.}) asserts that any problem that can be efficiently solved on \emph{any} reasonable machine model $\mathfrak{M}$ (von Neumann architecture, Harvard architecture, or whatever) can be efficiently solved on a probabilistic Turing machine, or more formally:
\begin{align}
  \label{e:universality}
  \bigcup\mbox{Poly}_{\mathfrak{M}} \;=\; \mathbf{BPP}.
\end{align}
In other words, the thesis implies that the set of problems solvable in polynomial time does not grow beyond \textbf{BPP} if we allow ourselves to vary the underlying model. Assuming the thesis is true, we do not need to worry about what model an algorithm is implemented on when discussing the computational complexity of various problems; we can simply focus on the abstract probabilistic Turing machine model and carry out our analyses in relation to it.\footnote{\label{fn:bpp-equals-p}We could have also expressed the thesis in terms of \textbf{P} rather than \textbf{BPP}. Although it was thought, for many years, that there are more problems efficiently solvable on a probabilistic Turing machine than on a standard Turing machine, a number of recent results have pointed in the opposite direction \citep[e.g.,][]{agrawal2004}, and it is now generally believed that classical probabilistic computation does not offer any performance advantage over classical deterministic computation \citep[Ch. 20]{arora2009}. In other words it is now widely believed that $\mathbf{P} = \mathbf{BPP},$ or that it is just as easy to solve a given problem on a deterministic Turing machine as it is on a probabilistic one. We have nevertheless stated the universality thesis in terms of \textbf{BPP} because this will prove convenient when it comes time to discuss the differences between classical and quantum computation. A quantum computer is, from one point of view, just another kind of probabilistic computer (that calculates probabilities differently), and it has the same success criterion as a classical probabilistic computer, i.e., we only demand that a given `solution' be correct with `high enough' probability.}

The second thesis we will introduce is called the \emph{invariance thesis} \citep[p. 5]{boas1990},\footnote{See also: \citet[p. 33]{goldreich2008}, who names it differently.} which asserts that given any two reasonable machine models $\mathfrak{M}_i$ and $\mathfrak{M}_j$, $\mathfrak{M}_i$ can efficiently simulate $\mathfrak{M}_j$; i.e. $\mathfrak{M}_i$ can solve any problem just as easily (in the sense explained above) as $\mathfrak{M}_j$ can. More formally:
\begin{align}
\label{e:invariance}
\forall_{i,j}~\mathfrak{M}_i \overset{poly}{\sim} \mathfrak{M}_j.
\end{align}
Note that the invariance thesis implies the universality thesis, but not vice versa.

\subsection{Physical perspectives on computer science}
\label{s:pccp0}

Neither the universality of Turing efficiency thesis, nor the invariance thesis, is a mathematical theorem. These statements can be true or false, and for a long time they were thought to be true, for none of the reasonable (classical) universal models of computation that had been developed since the time of Turing were found to be more efficient than the Turing model by more than a polynomial factor, and all of them had been shown to be able to simulate one another efficiently \citep[]{boas1990}. Over the last three decades, however, evidence has been mounting against universality and invariance, primarily as a result of the advent of quantum computing \citep[chs. 10, 15]{aaronson2013}.

Quantum mechanics, as we will see in the next section, is an irreducibly probabilistic theory. Thus a quantum computer is a type of probabilistic machine. Analogously to the way we defined \textbf{BPP} as the class of problems solvable (in the probabilistic sense) in polynomial time on a probabilistic Turing machine, we can also define the complexity class \textbf{BQP} (for `bounded error quantum polynomial') as the class of problems solvable (in the same sense) in polynomial time on a quantum computer. It is easy for a quantum computer to simulate a probabilistic Turing machine. Thus:
\begin{align}
  \label{e:bpp-subseteq-bqp}
  \mathbf{BPP} \subseteq \mathbf{BQP},
\end{align}
i.e., the class of problems efficiently solvable on a quantum computer is at least as large as the class of problems efficiently solvable on a probabilistic Turing machine. What is still unknown, though it would be surprising if it were not the case, is whether
\begin{align}
  \label{e:bpp-subsetneq-bqp}
  \mathbf{BPP} \subsetneq \mathbf{BQP},
\end{align}
i.e., whether the class of problems efficiently solvable on a quantum computer (a physically realisable computational architecture) is \emph{strictly larger than} the class of problems efficiently solvable on a probabilistic Turing machine. Note that if Eq.\ \eqref{e:bpp-subsetneq-bqp} is true, then both the universality and invariance theses are false.

There are some authors who view the consequences of the falsification of the universality thesis, in particular, to be profound. \citet{bernstein1997}, for example, take it that computational complexity theory ``rests upon'' this thesis (p. 1411), and that the advent of quantum computers forces us to ``re-examine the foundations'' (p. 1412) of the theory. The sense in which complexity theory rests upon universality is expressed by \citet[]{nielsenChuang2000}, who write that the falsity of the thesis implies that complexity theory cannot achieve an ``elegant, model independent form'' (p. 140). For Amit \citet[]{hagar2007b}, the failure of model-independence shakes, not only the foundations of complexity theory, but certain views in the philosophy of mind that depend on the model-independence of computational kinds (pp. 244--245).

If we actually examine what the universality thesis is saying, however, then it is not really clear, at least not \emph{prima facie}, how it can ground the model-independence of complexity-theoretic concepts. The statement of the thesis is that any efficiently solvable problem is solvable efficiently by a probabilistic Turing machine. A probabilistic Turing machine is a particular model of computation, though. How can a thesis whose very definition makes reference to a particular computational model ground the model-independence of computational complexity theory? In fact there is a weak notion of model-independence being alluded to here. The point \cite[see][p. 140]{nielsenChuang2000} is that, for any assertion of the form: `problem $P$ is efficiently solvable under computational model $\mathfrak{M}$', the qualification `under computational model $\mathfrak{M}$' can \emph{always} (given the truth of the universality thesis) be substituted with `by a probabilistic Turing machine' without changing the truth value of that sentence. Further, to show that such a sentence is true in general, it suffices to show that it is true for a probabilistic Turing machine. Finally, because `by a probabilistic Turing machine' qualifies every such sentence we can leave it off and still expect to be understood. `Problem $P$ is efficiently solvable by a probabilistic Turing machine' is thus abbreviated to `problem $P$ is efficiently solvable'. That we can do this is not insignificant, but this is arguably not a particularly deep sense of model-independence \citep[see][\S 11.6]{cuffaro2018b}.

Far more interesting in relation to the question of model-independence is the invariance thesis. Unlike the universality thesis, model-independence is built right into the very statement of invariance. For after all, it amounts to the quite direct claim that the details of any particular reasonable machine model, since these can be efficiently simulated by any other reasonable model, are irrelevant for the purposes of characterising the complexity of a given problem.

Invariance, if taken to be true without qualification, clearly brings with it an absolute notion of model-independence (at least with respect to physically reasonable models). And if taken to be false (as it seems we should, given the existence of quantum computation) it clearly precludes such a notion. Arguably, however, what is (and always was) most valuable about the idea of invariance is not the absolute model-independence that is implied when it is taken to hold without qualification. As we will see later, the term `quantum computer' does not refer to some single particular model of computation but is, rather, an umbrella term for a number of distinct computational models,\footnote{For instance, the quantum Turing model \citep[]{deutsch1985}, the quantum circuit model \citep[]{deutsch1989}, the cluster-state model \citep[]{briegel2009}, the adiabatic model \citep[]{farhi2000}, and so on.} all of which have been shown to be computationally equivalent in the sense that they are all efficiently simulable by one another.\footnote{See, for instance, \citet[][]{aharanov2007, raussendorf2002, nishimura2009}.} In other words, what we have learned from the study of quantum computing is that, in addition to the existence of an equivalence class of reasonable classical models that satisfy the invariance thesis with respect to one another, there is a further class of reasonable computational models, based on the principles of quantum mechanics, that satisfy the invariance thesis with respect to one another. Thus there are \emph{two} distinct equivalence classes of physically reasonable computational models from the viewpoint of computational complexity theory. This is a discovery.

Invariance, thought of as a guiding rule or methodological principle, rather than as an absolute thesis, can be understood as grounding these investigations, and arguably this was the point all along \citep[][\S 11.6]{cuffaro2018b}. Through the search for equivalence classes we carve out the structure of the space of computational models, yielding a notion of \emph{relative} model-independence among the machine models comprising a particular equivalence class. And the fact that relative model-independence exists within the space of computational models at all arguably tells us something deep about how computer science connects up with the world, for the differences in computational power between these two reasonable classes of computational model are best understood by considering the \emph{physics} needed to describe them. We discussed the physics of classical computers in Section \ref{s:classical-states}. In the next section we turn to the physics of quantum computers.

\subsection{Quantum states and operations}
\label{s:quantum-states-and-operations}

We saw earlier that in classical mechanics, assigning a particular state, $\omega = (\vec{q}, \vec{p})$, to a system fixes the answer to every conceivable yes-or-no experimental question that we can ask about it in advance, irrespective of whether we actually ask any questions or not. And we saw how to arrange the possible questions we can ask about a system, and their possible answers, into a truth-table-like structure like the one given in Figure \ref{f:truth-table}. Note that specifying the values of the answers to the questions on the right-hand side of the double-vertical line in Figure \ref{f:truth-table} is another way of representing the state of a system like the one depicted in Figure \ref{f:switches}. We saw how one can manipulate such a system in ways that can be represented abstractly as logico-mathematical operations on binary digits, where the binary digits are abstract representations of the properties of the system's (two-level) subsystems.

\begin{figure}
  \begin{center}
    \includegraphics[scale=0.35]{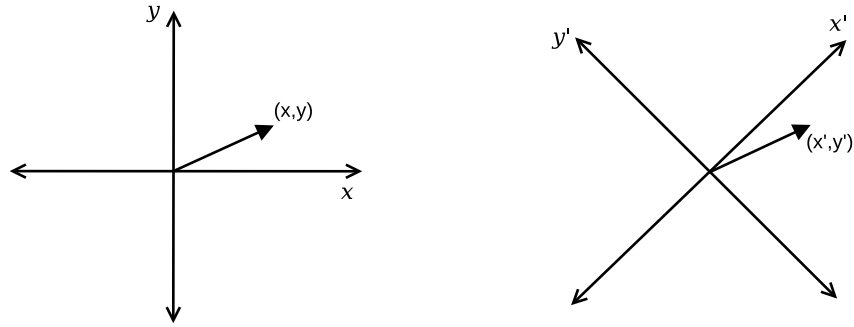}
  \end{center}
  \caption{An ordinary vector in the Cartesian plane. It can be decomposed into $x$ and $y$ coordinates, as on the left, or into the alternative coordinates $x'$ and $y'$ as on the right. Similarly, a state vector, in quantum mechanics, can be decomposed into more than one Hilbert space basis. Two common bases (but not the only two) with which to decompose the state vector for a qubit are the $\{| 0 \rangle, | 1 \rangle\}$ basis (known as the computational basis) and the $\{| + \rangle, | - \rangle\}$ basis.}
  \label{f:different-bases}
\end{figure}

The subject matter of \emph{quantum} mechanics, just as it is for classical mechanics, is physical systems, where these can be anything from a single particle to the entire physical universe. In other words, quantum mechanics is not just a theory of the small. It is, rather, a new universal language for describing physical systems,\footnote{See note \ref{fn:framework} above.} and it describes our experience with those systems better and more accurately than classical mechanics describes it. In quantum mechanics, just as in classical mechanics, a system can be, at a given moment in time, in any one of a number of possible physical states. These states are represented by \emph{state vectors} (or `wave functions'). The state vector for a two-dimensional quantum system or `qubit', for instance, is given in general by:
\begin{align}
  \label{eqn:general-form-state-vector}
  | \psi \rangle = \alpha| 0 \rangle + \beta| 1 \rangle.
\end{align}
When the complex numbers $\alpha$ and $\beta$ are both non-zero, a qubit is said to be in a \emph{superposition} of the \emph{basis vectors} $| 0 \rangle$ and $| 1 \rangle$. Just as an ordinary vector in the Cartesian plane can be decomposed, mathematically, into $x$ and $y$ coordinates, a state vector for a two-dimensional quantum system can be decomposed, mathematically, into two basis vectors for the (Hilbert) space that it is situated in (see Figure \ref{f:different-bases}). Two common bases (but not the only two) with which to decompose the state vector for a qubit are the $\{| 0 \rangle, | 1 \rangle\}$ basis (known as the computational basis) and the $\{| + \rangle, | - \rangle\}$ basis.

Associated with a given basis is a particular experimental question; for instance, ``Is the qubit in the state $| 0 \rangle$ (as opposed to the state $| 1 \rangle$)?'' in the case of the computational basis, and in the case of the $\{| + \rangle, | - \rangle\}$ basis, ``Is the qubit in the state $| + \rangle$ (as opposed to the state $| - \rangle$)?'' When a system in the state $\alpha| 0 \rangle + \beta| 1 \rangle$ is `asked' whether it is in state $| 0 \rangle$ as opposed to the state $| 1 \rangle$---i.e., when it is \emph{measured} in the computational basis----then with probability $|\alpha|^2$ the answer will come back as $| 0 \rangle$ and with probability $|\beta|^2$ it will come back as $| 1 \rangle$; i.e.,
\begin{align}
  \label{eqn:born-rule}
  M\big(\alpha| 0 \rangle + \beta| 1 \rangle\big) \;& \xrightarrow{\mathrm{Pr} \;=\; |\alpha|^2} \;  | 0 \rangle \\
  M\big(\alpha| 0 \rangle + \beta| 1 \rangle\big) \;& \xrightarrow{\mathrm{Pr} \;=\; |\beta|^2} \; | 1 \rangle
\end{align}
where $M$ is a \emph{measurement operator} used to represent a computational basis measurement. Note that there is also a basis corresponding to the question: ``Is the qubit in the state $| \psi \rangle$ (as opposed to the state $| \phi \rangle$)?'' Why not just ask this question? One reason is that measuring a quantum-mechanical system involves using a concrete physical device, and it turns out that it is technologically more feasible to construct measurement devices corresponding to some experimental questions than to others \citep[\S 6.5]{3m2020}.

In terms of their dynamics, states of quantum systems evolve linearly:
\begin{align}
  \label{eqn:linear}
  U\big(\alpha| 0 \rangle + \beta| 1 \rangle\big) = \alpha U| 0 \rangle + \beta U| 1 \rangle
\end{align}
and unitarily:
\begin{align}
  \label{eqn:unitary}
  UU^\dagger| \psi \rangle = U^\dagger U| \psi \rangle = | \psi \rangle
\end{align}
over time, where $U$ is a \emph{unitary operator} and $U^\dagger$ is its \emph{adjoint}.

Classical and quantum mechanics differ in the way that answers to experimental questions are determined. In contrast to classical mechanics, specifying a quantum system's state at a given moment in time does not fix in advance the answers to the experimental questions that can be asked of that system, or as \citet[]{bub-pitowsky2010} put it, the quantum state is not a \emph{truthmaker} in relation to those questions. It fails to be a truthmaker in two senses: First, a given state specification yields, in general, only the probability that the answer to a given experimental question will take on one or another value when asked. This in itself is not as much of a departure from classical mechanics as one might think, however, because conditional upon the selection of an experimental question, one can, in quantum mechanics, describe the observed probabilities as stemming from a prior classical probability distribution over the dynamical properties of the system that \emph{is} determined in advance by the quantum state.\footnote{This is analogous to the way we interpret probabilities in classical \emph{statistical} mechanics.}

This brings us to the second, more important, sense in which the state of a quantum-mechanical system fails to be a truthmaker in relation to questions that can be asked about that system's dynamical properties: The probability distributions associated with the answers to individual experimental questions cannot be embedded into a global prior probability distribution over all of the answers, as they can be for a classically describable system (see Figure \ref{f:truth-table}). In quantum mechanics one can only say that \emph{conditional} upon our inquiring about the observable $A$, there will be a particular probability for the answer to that question to take on a particular value. Thus in quantum mechanics, unlike in classical mechanics, we actually have to ask the system a question in order to get an answer from it.\footnote{See also \citet[chs. 1 and 6]{3m2020}, who call this (ironically) the `small' measurement problem in contrast to the `big' (this label is also ironic) problem described in the previous paragraph. These labels are originally due to \citet[]{bub-pitowsky2010}, who used them ironically too.}

When quantum-mechanical systems are combined, they can sometimes become \emph{entangled} with one another. Consider, by way of contrast, the simple illustration in Figure \ref{f:switches} of a classically describable system composed of many subsystems. This system is in a \emph{product state}, which means that the state of the overall system can be expressed as a product of the states of each individual subsystem, i.e., $0_A1_B1_C0_D$, where the individual states of the subsystems are $0_A$, $1_B$, $1_C$, and $0_D$. In classical mechanics there is no other way to describe the combined state of a number of subsystems. Even when these subsystems are correlated with one another (as the state of one side of a coin is correlated with the state of the other side, for instance), as long as we include facts about those correlations in our description of the overall system, then its overall state can be factored into the individual states of its subsystems \citep[]{bell1981}. This is true even if the subsystems happen to be far apart, for instance if I print out two copies of a document and mail each of them to distant parts of the globe.

In quantum mechanics, in contrast, a system composed of multiple subsystems can sometimes be in a state like the following:
\begin{align}
  \label{e:entanglement-comp}
  \frac{1}{\sqrt 2}\big(| 0 \rangle_A| 1 \rangle_B \;-\; | 1 \rangle_A| 0 \rangle_B\big).
\end{align}
This state, describing a system composed of two subsystems $A$ and $B$, is \emph{not} a product of the individual states of its subsystems, for there is no way to factorise it into a state of the form: $| \psi \rangle_A| \phi \rangle_B$ (the reader is invited to try). It is sometimes said that in an entangled quantum system the whole is somehow prior to the parts \citep[]{howard1989}, or that the systems are \emph{nonlocally correlated} \citep[]{maudlin2011}, for if Alice measures her qubit in the computational basis and receives the result $| 0 \rangle_A$, then a measurement by Bob on his qubit (which in principle he may have taken to the opposite side of the universe) is thereby instantaneously determined to be $| 1 \rangle_B$, although there is no way for Bob to take advantage of this instantaneous determination for the purpose of predicting the outcome of a measurement on his subsystem, short of waiting for Alice to send him a classical signal (a text message or a phone call, for instance).\footnote{This restriction on the use that can be made of nonlocal correlations in quantum mechanics is called `no signalling'. For discussion, see \citet{bub2016, cuffaroInfCausality}.}

Note that the correlation between the results of computational (i.e., $\{| 0 \rangle, | 1 \rangle\}$) basis measurements on an entangled system in this state do not, in themselves, represent as much of a departure as one might think from what one can describe using classical language. For with respect to this specific correlation, \citet{bell1964} showed that it can be seen as arising from a classically describable system that has been prepared in a particular way. It is only when we consider other experiments on such a system that we see a departure from the possibilities inherent in classical description. In particular, because
\begin{align}
  \label{e:plus-zero}
  | + \rangle = \frac{| 0 \rangle + | 1 \rangle}{\sqrt 2}, \qquad | - \rangle = \frac{| 0 \rangle - | 1 \rangle}{\sqrt 2},
\end{align}
Eq.\ \eqref{e:entanglement-comp} can be rewritten, in the $\{| + \rangle, | - \rangle\}$ basis, as:
\begin{align}
  \label{e:entanglement-x}
  \frac{1}{\sqrt 2}\big(| + \rangle_A| - \rangle_B \;-\; | - \rangle_A| + \rangle_B\big).
\end{align}
Indeed, we obtain the \emph{same form} for the state of the system \emph{regardless} of which basis we choose to express it in:
\begin{align}
  \label{e:entanglement-arbitrary}
  \frac{1}{\sqrt 2}\big(| b_1 \rangle_A| b_2 \rangle_B \;-\; | b_2 \rangle_A| b_1 \rangle_B\big).
\end{align}
Eqs.\ \eqref{e:entanglement-comp}, \eqref{e:entanglement-x}, and \eqref{e:entanglement-arbitrary} all yield the same probabilities for the results of experiments.

If we collect a number of pairs of subsystems prepared in this entangled state, and then ask of each pair: ``Is this pair in the state $| 0 \rangle| 1 \rangle$ (as opposed to $| 1 \rangle| 0 \rangle$)?,'' then we will find that the answers `yes' and `no' will be obtained with equal probability. Conditional upon this question, we can imagine these answers as arising from a prior classical probability distribution over the properties of the subsystem pairs, half of which we imagine to be in the state $| 0 \rangle| 1 \rangle$, and half of which we imagine to be in the state $| 1 \rangle| 0 \rangle$. This prior probability distribution is \emph{incompatible}, however, with the prior classical probability distribution that we might imagine to be responsible for the answers yielded from repeatedly asking the question: ``Is this pair in the state $| + \rangle| - \rangle$ (as opposed to $| - \rangle| + \rangle$)?'' In the context of the latter question, we can imagine the answers as arising from a prior classical probability distribution over the properties of the subsystem pairs, half of which we imagine to be in the state $| + \rangle| - \rangle$, and half of which we imagine to be in the state $| - \rangle| + \rangle$. Asking a question of an imagined ensemble of systems, half of which are in the state $| 0 \rangle| 1 \rangle$, and half of which are in the state $| 1 \rangle| 0 \rangle$, will yield different statistics, however, than the ones that would be yielded by asking the same question of an imagined ensemble of systems, half of which are in the state $| + \rangle| - \rangle$, and half of which are in the state $| - \rangle| + \rangle$. These probability distributions are, in this sense, incompatible. And yet \emph{all} of these statistics are predicted by one and the same quantum state, the one expressed in the unitarily equivalent forms given in Eqs.\ (\ref{e:entanglement-comp}--\ref{e:entanglement-arbitrary}). In quantum mechanics, unlike in classical mechanics, the questions we ask or do not ask actually matter for the ways that we can conceive of the underlying properties of a physical system, even when the state of that system is fully specified \cite[\S 6.5]{3m2020}.

\section{Quantum computation and parallel universes}
\label{s:many-worlds}

The are problems that are hard for a classical computer, but that a quantum computer can solve easily (where `easy' and `hard' are meant in the computational sense defined above in Section \ref{s:classical-computers}).\footnote{There are also problems for which a quantum computer, despite being unable to solve them easily, can nevertheless solve them significantly \emph{more} easily than a classical computer can. An example is the problem to search an unstructured database, for which a quantum (``Grover's'') algorithm can reduce the number of steps required by a quadratic factor over any known classical algorithm. See: \citet[]{grover1996}, \citet[]{bennett1997}, and for further discussion see \citet[p. 269]{cuffaro2018b}.} The most famous example of such a problem is that of factoring large integers into primes. The best-known classical algorithm for factoring is the number field sieve \citep[]{lenstra1990}, which takes on the order of $2^{(\log N)^{1/3}}$ steps to factor a given integer $N$.\footnote{For the meaning of `on the order of' see fn. \ref{fn:big-oh}.} No one has yet proven that this is the best that a classical computer can do. All the same, encryption algorithms such as RSA (named for the paper by \citeauthor*[]{rsa1978} in which it was introduced), that are today widely used for secure transactions on the internet, rely on the assumption that it is; i.e., that one cannot factor in polynomial time. In \citeyear{shor1994}, Peter Shor discovered a \emph{quantum} algorithm---now known as Shor's algorithm---that can factor integers in on the order of $\log N$ steps for a given integer $N$---an exponential `speedup' over the number field sieve.

What explains this phenomenon? What is it, exactly, that allows quantum computers to compute certain problems more efficiently than classical computers can? Surprising as it may seem, there is still no consensus regarding the answer, neither among the researchers working in the field, nor among the philosophers commenting on this work. By contrast, in the popular literature on the topic, one answer has tended to dominate all the others. This is the idea that quantum computers are actually \emph{parallel} devices that perform exponentially many classical computations simultaneously, with each computation taking place in a parallel physical universe (or `world'), different in certain ways from our own but just as real as it is.

Perhaps the strongest proponent of this view, which we will call the \emph{many-worlds explanation} of the power of quantum computers, is one of the fathers of the field, David Deutsch, discoverer of the first quantum algorithm and of the universal quantum Turing machine \citep[]{deutsch1985}. The many-worlds explanation of quantum computing is, for Deutsch, not just a speculative conjecture. For Deutsch, many worlds are the only plausible explanation for how quantum computers work. As he puts it in his 1997 book, \emph{The Fabric of Reality}: ``[t]o those who still cling to a single-universe world-view, I issue this challenge: Explain how Shor's algorithm works'' \citeyearpar[p. 217]{deutsch1997}.

Before we discuss what those who, like Deutsch, defend this view of quantum computation are really saying, let us review briefly what the view amounts to in more general terms in the context of quantum mechanics, for what the many-worlds explanation of quantum computing amounts to is an application of this more general philosophical interpretation of quantum mechanics to the case of quantum-mechanical computers.\footnote{The interpretation of quantum mechanics that we will be discussing in this section is one of a number of related interpretations of quantum mechanics that are collectively referred to as the `Everett interpretation'. These include but are not limited to Hugh Everett III's original formulation \citep[]{everett1956}, the `Berlin Everettianism' of Christoph \citet[]{lehner1997}, Lev Vaidman's version of Everett \citep[]{vaidman1998}, so-called `many minds' variants \citep[]{albert1988}, and finally the `many-worlds' variants that are the direct inspiration for the many-worlds explanation of quantum computing. Belonging to the last group are Bryce DeWitt's \citeyearpar{dewitt1971} view, as well as the `Oxford Everett' interpretation \citep[]{deutsch1997, saunders1995, wallace2003, wallace2012} with which we will be mostly concerned here.} Thus far, our examples of quantum-mechanical systems have been of qubits, either considered singly or in combination with other qubits. But quantum mechanics is not restricted, of course, to describing qubits. Quantum mechanics is, rather, a universal language for describing any physical system, be it a single particle or (in principle) the entire physical universe. From the point of view of quantum mechanics, you and I, the laptop I am writing this chapter with, the table it is sitting on, and everything else in the world is a quantum-mechanical system. And just as a qubit can be in a superposition of its basis states (see Eq.\ \eqref{eqn:general-form-state-vector} and Figure \ref{f:different-bases}), so too can \emph{any} physical system. As Schr\"odinger famously remarked, according to quantum mechanics there are even ways of preparing a cat in a superposition of `dead' and `alive' states \citep[]{schrodinger1935a}.

When we actually come to pose a question to the system; i.e., to measure the qubit, the cat, or what have you, we never actually find that it is in a superposition state. Quantum mechanics predicts that the cat will be found to be either dead or alive, the qubit to be either $|0\rangle$ or $|1\rangle$, and so on, with a certain probability. Thus, unlike the physical state of a system as described by classical mechanics, we cannot simply read off from the quantum state description of a system what the outcome of an experiment on that system will be (except in special cases). This raises the question of how to interpret the quantum state. Are we to take superpositions literally, despite the fact that we never observe them? What would it mean if we did? The proponent of the many-worlds interpretation of quantum mechanics answers yes to the first question. As for the second question, the view of the many-worlds advocate is that each branch of a superposition represents a distinct, classically describable (from the point of view of that branch), physical universe.

Just as it is with a cat, or with any physical system, so it is with a quantum computer. What is different, arguably, about a quantum computer (according to the defender of the many-worlds explanation of quantum computing) is that in a quantum computer the resources made available by these multiple physical universes are harnessed to perform computations. This is a striking claim. But it can be said, in favour of the many-worlds explanation, that some quantum algorithms certainly do give us this impression. The following evolution is representative of a typical step in many quantum algorithms (note that normalisation factors have been omitted for simplicity):
\begin{align}
\label{e:parallel}
\sum_{x=0}^{2^n-1} | x \rangle^n | 0 \rangle \rightarrow \sum_{x=0}^{2^n-1} | x \rangle^n | f(x) \rangle,
\end{align}
What is being depicted here is the state of a quantum computer, composed of $n+1$ qubits, where the first $n$ qubits are called the \emph{input qubits} of the computer, and the last qubit is called the computer's \emph{output qubit}. The computer begins in the state described on the left-hand side of the equation, and then transitions to the state described on the right-hand side.

This notation is very compact, so let us unpack it. Considering, first, the left-hand side: If we take the $n$ input qubits as comprising, together, one subsystem of the system, and the output qubit as comprising another subsystem, then we can say that the input and output subsystems of the computer begin in a product state. In other words we can write the left-hand side of the equation in product form as:
\begin{align}
  \label{e:product}
  {\rm LHS} \;=\; \left(\sum_{x=0}^{2^n-1} | x \rangle^n\right)| 0 \rangle.
\end{align}
Note that, in Eq.\ \eqref{e:product}, the $n$ input qubits are together in a superposition of all of the possible computational basis states for a system of $n$ qubits. This is obscured somewhat, both by the summation notation as well as by the shorthand, $| x \rangle^n$, being used to represent each superposition term. To clarify this, recall that for a two-dimensional system, i.e., a single qubit, the computational basis states are, as we pointed out earlier (see Eq.\ \eqref{eqn:general-form-state-vector} and Figure \ref{f:different-bases}), the states $| 0 \rangle$ and $| 1 \rangle$. As for a system composed of \emph{two} qubits, it has four computational basis states, namely, $| 0 \rangle| 0 \rangle$, $| 0 \rangle| 1 \rangle$, $| 1 \rangle| 0 \rangle$, and $| 1 \rangle| 1 \rangle$. In general, for a system of $n$ qubits there are $2^n$ basis states:
\begin{align}
\label{e:basis-n-dim}
  | 0 \rangle| 0 \rangle \dots | 0 \rangle| 0 \rangle, \nonumber \\
  | 0 \rangle| 0 \rangle \dots | 0 \rangle| 1 \rangle, \nonumber \\
  | 0 \rangle| 0 \rangle \dots | 1 \rangle| 0 \rangle, \nonumber \\
  \dots \nonumber \\
  | 1 \rangle| 1 \rangle \dots | 1 \rangle| 1 \rangle,
\end{align}
each of which can be thought of as representing a binary number. For short, we can represent these in decimal as $\{| \mathbf{0} \rangle, | \mathbf{1} \rangle, | \mathbf{2} \rangle, | \mathbf{3} \rangle,\; \dots \; | \mathbf{2^{n-1}} \rangle\}.$ A superposition of all of these basis states is then given by:
\begin{align}
  \label{e:superpos}
  | \mathbf{0} \rangle \;+\; | \mathbf{1} \rangle \;+\; | \mathbf{2} \rangle \;+\; | \mathbf{3} \rangle \;+\; \dots \;+\; | \mathbf{2^{n-1}} \rangle \;=\; \sum_{x=0}^{2^n-1} | x \rangle^n,
\end{align}
exactly as we see in Eq.\ \eqref{e:product}.

On the right-hand side of Eq.\ \eqref{e:parallel}, the $n+1$ qubits of the quantum computer are no longer in a product state. The state of the computer has now transitioned to an \emph{entangled} state: a superposition in which the state obtained for the output qubit, conditional upon measuring it, is correlated with the state that will be obtained for the input qubits, conditional upon measuring them. For instance, if we measure the input qubits and get the state $| \mathbf{1} \rangle$, then the state of the output qubit will be $| f(\mathbf{1}) \rangle$, for the given function $f$, and similarly for every other value of $x$. What is the function $f$? That will depend on the particular problem that the algorithm is designed to solve. More importantly, from the point of view of our current discussion, notice that the right-hand side of the equation encodes exponentially many evaluations of that function in the state of the computer! We can, or so it seems, appeal to the results of all of these evaluations to help us with whatever problem we are trying to solve.

In reality things are not so easy, for when we actually come to read off the result of the computation, we will only ever find the computer's output qubit to be in \emph{one} of its exponentially many superposition terms; i.e., in some state $|f(a)\rangle$ for some one particular $a$ in the domain of $f$. If we are tempted to read the presence of these exponentially many function evaluations in the description of the state of the computer literally, then the fact that only one of them can ever be accessed in a particular run should give us some pause \citep[cf.][p. 38]{mermin2007}. That said, in any given run, there will be an, in general, non-zero probability of obtaining any one of them, and this, perhaps, speaks in favour of viewing them all as somehow literally there. As for the goal (which we should not lose sight of) of actually solving the problem at hand, what is just as important as achieving the form of the state of the computer on the right-hand side of Eq.\ \eqref{e:parallel} is the \emph{next} step in a computation like this one, which requires that we manipulate the system cleverly enough so that one of the desired solutions is found with higher probability than the other, undesirable, solutions \citep[\S 3]{pitowsky2002}.

The many-worlds explanation of a quantum computational process enjoins us to take an evolution like the one given in Eq.\ \eqref{e:parallel} ontologically seriously. It affirms, in other words, that things are indeed as they seem: The computer most definitely \emph{is} performing many simultaneous function evaluations in parallel when it is in a state like the one on the right-hand side of Eq.\ \eqref{e:parallel}. Moreover the many-worlds explanation directly answers the question of \emph{where} this parallel computation occurs, namely in distinct physical universes. For this reason it is also, arguably, one of the more intuitive of the physical explanations of quantum speedup. And it is certainly thought-provoking. It is no wonder, then, that is the one most often mentioned in the popular literature on quantum computation. And it has advocates in the more serious literature as well.\footnote{In addition to Deutsch's 1997 book, see \citet[]{deutsch2010}, and see also \citet[\S 7]{vaidman2018} and \citet[Ch. 10]{wallace2012}. The strongest and most in-depth defence of the many-worlds explanation of quantum computing that I am aware of is the one given by \citet[]{hewittHorsman2009}.}

So far we have highlighted one potential problem for the many-worlds explanation: Despite the appearance of parallel processing in an equation like Eq.\ \eqref{e:parallel}, only one of these function evaluations is ever accessible on a given run of the computer. This is not to say that no story can be told from the point of view of the many-worlds explanation about why this is so, but at the very least this is enough to cast doubt on the claim that an evolution like the one given in Eq.\ \eqref{e:parallel} constitutes, all by itself, evidence for the many-worlds explanation of the power of quantum computers. We will return to this issue later. But for now we need to consider a somewhat deeper problem faced by the advocate of the many-worlds explanation. Recall that decomposing a qubit in the state $| \psi \rangle$ into the computational basis states $| 0 \rangle$ and $| 1 \rangle$ (see Eq.\ \eqref{eqn:general-form-state-vector} and Figure \ref{f:different-bases}) is only one way to express this state. The same state can also be expressed in other bases besides the computational basis. This means that a system that is in the state
\begin{align}
  \label{e:qubit-superpos}
\frac{1}{\sqrt 2}| 0 \rangle + \frac{1}{\sqrt 2}| 1 \rangle,
\end{align}
from the point of view of the computational basis, is simply in the state:
\begin{align}
| + \rangle
\end{align}
from the point of view of the $\{| + \rangle, | - \rangle\}$ basis (see Eq.\ \eqref{e:plus-zero}). But now it is no longer clear just how many universes we should take this qubit to exist in. If we decompose its state in the computational basis then it would seem that it exists in two worlds, and if we decompose it into the $\{| + \rangle, | - \rangle\}$ basis then it would seem that it only exists in one. Which is it? If we decide that one of these decompositions is to be preferred to the other, then the challenge for the advocate of the many-worlds explanation is to give a compelling reason why, for \emph{prima facie} there does not appear to be any reason for preferring one of these decompositions over the other.

This is known as the \emph{preferred basis problem}, and it is a problem for the many-worlds view more generally, i.e., not just in the context of quantum computing. In this more general context, advocates of the many-worlds picture \citep[see, for instance,][]{wallace2003} attempt to solve the preferred basis problem by appealing to the dynamical process of \emph{decoherence}. In quantum mechanics, as we saw (see Eq.\ \eqref{eqn:general-form-state-vector} and Figure \ref{f:different-bases}), the state of a system at a particular moment in time is described by a state vector that evolves linearly and unitarily. However the evolution thereby described is the evolution of a system that is completely isolated from its external environment, which is an idealisation; in reality it is actually never possible to completely isolate a system from its environment,\footnote{At the very least, the gravitational effects of other distant systems will not be able to be neglected.} unless, perhaps, we take our system of interest to be the universe in its entirety.\footnote{Some philosophers have questioned whether we should think of even the universe as a whole as a closed system (see, for instance, \citeauthor{cuffaroHartmannOpenSystems}, \citeyear{cuffaroHartmannOpenSystems}; \citeauthor{wallace2012}, \citeyear[\S 10.5]{wallace2012}).} But barring the universe as a whole, when a system interacts with an external environment---for example with our measurement apparatus as we ask the system an experimental question---then the terms in the superposition describing its state begin to decohere \citep[]{zurek2003} and come to achieve a kind of independence from one another (although they never decohere completely). On the many-worlds picture we are to think of such (approximately) decoherent terms as existing in independently evolving physical universes. And in each of these independently evolving physical universes there is in addition a different version of ourselves, all of whom receive a different answer to the experimental question that was asked. And with every additional question, the universe as we know it is branching, spawning exponentially more and more versions of the system, and more and more versions of ourselves along with it, and so on and on and on.

Fundamentally, however, decoherence is an approximate phenomenon, and some small amount of residual interference between worlds always remains despite it. Nevertheless decoherence tells us that, when the environment and system happen to be correlated in a particular way, then \emph{a particular basis will emerge with respect to which} we can effectively describe the superposition terms expressed in the state of the system as evolving independently of one another. As David \citet[p. 90]{wallace2003} puts it: ``the basic idea is that dynamical processes cause a preferred basis to emerge rather than having to be specified a priori.'' In this way, superposition terms that for all practical purposes evolve stably and independently over time with respect to the decoherence basis can be identified with different copies of measurement pointers, cats, experimenters, and whatever else is theoretically useful for us to include in our ontology: ``the existence of a pattern as a real thing depends on the usefulness---in particular, the explanatory power and predictive reliability---of theories which admit that pattern in their ontology'' \citep[p. 93]{wallace2003}. Whatever else one may think of the many-worlds view, decoherence, at least, does provide a principled way to identify worlds in the wave-function.

For the advocate of the many-worlds explanation of \emph{quantum computation}, however, there is still a problem. Appealing to decoherence may solve the preferred basis problem for the purposes of describing the world of our everyday experience,\footnote{The preferred basis problem is not the only challenge that needs to be met by an advocate of the Everett interpretation of quantum mechanics. Another issue that has been much discussed in recent literature is the problem of how to account for probabilities on the Everettian view. For more on this issue see \citet[]{adlam2014}, \citet[]{dawidThebault2015}, \citet[]{greavesMyrvold2010}, \citet[]{vaidman1998, vaidman2012}, and \citet[]{wallace2007}.} but the inner workings of a quantum computer are not part of that everyday experience. The problem is not just that qubits are too small to see. The problem is that the superpositions characteristic of quantum algorithms are \emph{coherent} superpositions (\citealt[p. 278]{nielsenChuang2000}; see also \citealt{aaronson2013b, cuffaro2012}). Thus the terms in the wave-function of a quantum computer do not seem to meet the criterion for world-identification advocated for by the many-worlds view.

Now, to be fair, a similar thing can be said with regard to our everyday experience. So-called decoherent superpositions are (as I mentioned) not \emph{really} decoherent, after all. But they are decoherent enough, and for long enough, that it is useful (or so says the many-worlds advocate) to think of the terms in such a superposition as independent, and the worlds that they describe as ontologically real. Likewise, although the superposition on the right-hand side of Eq.\ \eqref{e:parallel} is actually a coherent superposition, it may nevertheless be useful to think of the terms in that superposition as independent, at least for the short time that the quantum computer is in that state \citep[p. 876]{hewittHorsman2009}. The problem, however, is that it is the very fact that they are coherent that allows us to `cleverly' extract desirable solutions from these superpositions with higher probability than undesirable solutions \citep[]{bub2010, duwell2007}.

Even if we grant that there is some heuristic value in describing a quantum computer in a superposition state as evaluating functions in exponentially many parallel worlds (I do not doubt that this was of some heuristic use to Deutsch, for instance, even if I question whether it is \emph{necessary} to think of such superpositions in this way), it does not follow that this is enough to licence granting ontological status to those worlds. Wallace \citeyearpar[p. 93]{wallace2003} mentions (as we saw) explanatory power and predictive reliability, for instance, and discusses the way that these and other ideas are applied in contemporary physics to support the many-worlds view outside of the context of quantum computing. It is not at all clear that these criteria are met in the context of quantum computing, however, and even Wallace admits that they for the most part are not: ``There is no particular reason to assume that \emph{all} or even \emph{most} interesting quantum algorithms operate by any sort of `quantum parallelism''' \citep[p. 70, n. 17]{wallace2010}. Wallace goes on: ``Shor's algorithm, at least, does seem to operate in this way'' (ibid.), but he does not describe how. Yet there are very plausible accounts of how Shor's algorithm works that do not appeal to massive parallelism at all \citep[see][]{bub2010}. Far from it, on Jeffrey Bub's account of Shor's algorithm, the quantum algorithm is more efficient than known classical algorithms because it performs \emph{fewer}, not more, computations \citep[see also][]{bub2008}.

The final reason that I will mention for being skeptical of the many-worlds explanation of quantum computing is that it only really seems to be useful in the context of one particular model of quantum computing. This is the so-called \emph{quantum circuit model}, the model for which many of the first quantum algorithms were designed \citep[]{deutsch1989}. This model is useful for abstract theoretical purposes, as well as for pedagogical purposes, as it borrows many familiar ideas from the classical circuit model of computation (see Figure \ref{f:classical-gates}). In the quantum circuit model, similarly to the classical circuit model, logical circuits are constructed out of various `quantum logic gates'. These instantiate unitary transformations of one or more qubits that are prepared beforehand in computational basis states (typically qubits begin a computation in the state $| 0 \rangle$). The unitary gates transform the qubits' states into various superpositions of computational basis states, and at the end of the computation a measurement is performed, again in the computational basis, on (some of) the qubits and the results are read out.

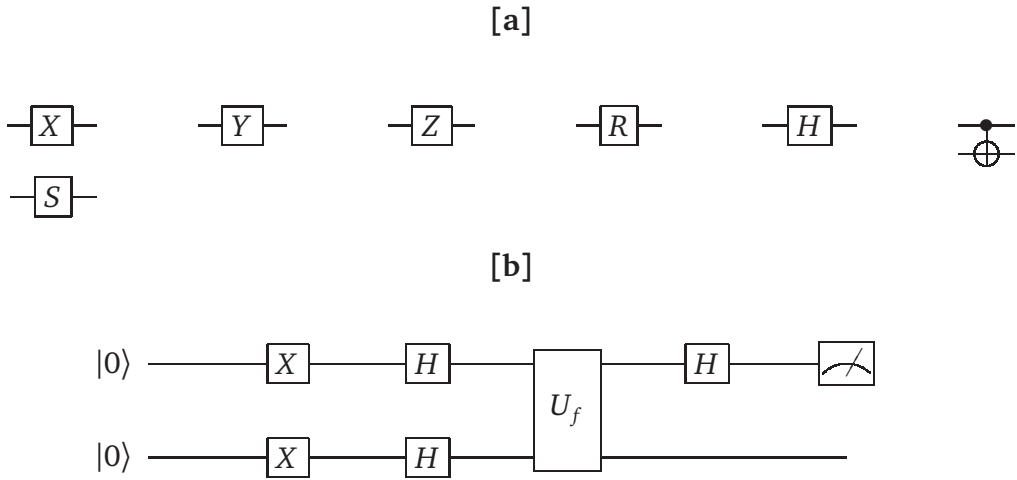
\begin{figure}
  \begin{center}\textbf{[a]}\end{center}
  \begin{align*}
    \xgate & & \ygate & & \zgate & & \rphasegate & & \hgate & & \cnotgate \\
    \sphasegate
  \end{align*}
  \begin{center}\textbf{[b]}\end{center}
  \begin{align*}
    \Qcircuit @C=.1em @R=.91em @! {
    \lstick{| 0 \rangle} & \qw & \gate{X} & \qw & \gate{H} & \qw & \multigate{1}{U_f} & \qw & \gate{H} & \qw & \meter \\
    \lstick{| 0 \rangle} & \qw & \gate{X} & \qw & \gate{H} & \qw & \ghost{U_f}        & \qw & \qw      & \qw & \qw \\
    }
  \end{align*}
  \caption{\textbf{[a]} A number of important one- and two-qubit quantum gates. The $X$, $Y$, $Z$, $R$, $H$, and ${\rm CNOT}$ gates together form the \emph{Clifford group}. If we add the $S$ gate to the Clifford group, they together form a \emph{universal set}. \textbf{[b]} A quantum circuit diagram for Deutsch's Algorithm, which determines whether a given function $f$ on one bit is constant ($f(0) = f(1)$) or balanced ($f(0) \not= f(1)$).}
  \label{f:quantum-circuit-model}
\end{figure}

Figure \ref{f:quantum-circuit-model}[a] depicts a number of important one- and two-qubit quantum gates. The $X$ gate implements a qubit-flip operation; i.e., it takes $| 0 \rangle \to | 1 \rangle$ and vice versa. The $Y$ gate takes $| 0 \rangle \to i| 1 \rangle$ and $| 1 \rangle \to -i| 0 \rangle$. The $Z$ gate takes $| 0 \rangle \to | 0 \rangle$ and $| 1 \rangle \to -| 1 \rangle$. The $R$ gate takes $| 0 \rangle \to | 0 \rangle$ and $| 1 \rangle \to i| 1 \rangle$. The $H$ (or Hadamard) gate takes $| 0 \rangle \to \nicefrac{(| 0 \rangle \,+\, | 1 \rangle)}{\sqrt 2}$ and $| 1 \rangle \to \nicefrac{(| 0 \rangle \,-\, | 1 \rangle)}{\sqrt 2}$. The $S$ gate takes $| 0 \rangle \to | 0 \rangle$ and $| 1 \rangle \to e^{\nicefrac{i\pi}{4}}| 1 \rangle$. At the extreme right is the two-qubit ${\rm CNOT}$ (or controlled-not) gate. It leaves the topmost qubit unchanged. The bottom qubit is then assigned the output of taking the exclusive-or of both, i.e., this gate takes $| 0 \rangle| 0 \rangle \to | 0 \rangle| 0 \rangle$, $| 0 \rangle| 1 \rangle \to | 0 \rangle| 1 \rangle$, $| 1 \rangle| 0 \rangle \to | 1 \rangle| 1 \rangle$, and $| 1 \rangle| 1 \rangle \to | 1 \rangle| 0 \rangle$. The $X$, $Y$, $Z$, $R$, $H$, and ${\rm CNOT}$ gates together form the \emph{Clifford group} of gates, which we will have more to say about later. If we add the $S$ gate to the Clifford group, they together form a \emph{universal set} of gates, i.e., any quantum circuit implementing any series of unitary transformations can be simulated to arbitrary accuracy using combinations of these seven gates.

Figure \ref{f:quantum-circuit-model}[b] depicts a quantum circuit diagram for Deutsch's Algorithm, which determines whether a given function $f$ on one bit is constant ($f(0) = f(1)$) or balanced ($f(0) \not= f(1)$): Two qubits are prepared in the product state $| 0 \rangle| 0 \rangle$ and are each sent through an $X$-gate and a Hadamard gate, after which they are together input to the two-qubit entangling unitary gate $U_f$. The first qubit is then sent through a further Hadamard gate and finally measured to yield the answer \citep[see][]{deutsch1989}.

Just as there are various models of universal classical computation (for instance the various versions of the Turing machine, as well as the von Neumann architecture, and so on, that I mentioned above), there are various models of universal quantum computation. One computational model that presents a particularly difficult problem for those who would advocate for the many-worlds explanation is the \emph{cluster-state} model of quantum computing, also known as \emph{measurement-based} and \emph{one-way} quantum computing \citep[]{raussendorf2002, raussendorf2003, nielsen2006}.\footnote{For introductions to cluster-state quantum computing aimed at philosophers, see \citet[\S 4]{cuffaro2012} and \citet[\S 4.5]{duwell2018}.} In the cluster-state model, the computer begins, not in a product state but in a highly entangled state, and measurements are performed not only at the end of the computation but throughout it. These measurements are \emph{adaptive}, in the sense that each measurement is performed in a \emph{different basis}, which depends on the random outcomes of whatever previous measurements have been performed.

Why is this a problem for the advocate of the many-worlds explanation of quantum computing? Because the fact that measurements are adaptive means, on the one hand, that there is no principled way to select a preferred basis \emph{a priori} in the context of a given computation \citep[\S 4]{cuffaro2012}. Whichever basis we choose, few qubits will actually be measured in that basis in the course of the computation. On the other hand, there is no sense in which we can say that a preferred basis `emerges' from the computational process. There is therefore no way to identify the worlds that the computation as a whole is supposed to be happening in.

As I alluded to above, both the cluster-state model and the circuit-model are universal models of quantum computing. Thus anything that the circuit model can do can be done (and, as it happens, just as efficiently) in the cluster-state model and vice versa. Perhaps, then, it could be argued that understanding the `true nature' of algorithms in the cluster-state model requires that we first translate them into the language of the circuit model, though I can think of no reason why one should think so other than a desire to hold on to the many-worlds idea come what may. The proper response to anyone who would put forward such an argument is that a large part of what motivates those who adhere to the many-worlds explanation of quantum computing in the first place, is that it is useful for algorithm analysis and design to believe that a quantum computer is carrying out its computations in parallel worlds. This does not seem to be so for the cluster-state model. On the contrary, dogmatically holding on to the view that many worlds are, at root, physically responsible for the speedup evinced in the cluster-state model is at best useless, for it is of no help in designing algorithms for the cluster-state model. At worst, dogmatically holding on to the many-worlds idea could prove positively detrimental if it prevents us from exploiting the power of the cluster-state model or discovering other quantum computational models in the future.

\section{Quantum computation and entanglement}
\label{s:entanglement}

The many-worlds explanation of the power of quantum computers is not the only explanation that has been proposed by philosophers. I have already mentioned the upshot of Bub's analysis of Shor's algorithm in the previous section. Bub's view, more generally \citeyearpar[]{bub2006, bub2010}, is that the key to understanding the power of quantum computers lies in understanding the way that they exploit the novel logico-probabilistic structure of quantum mechanics \citep[]{pitowsky1989}. We saw above that in quantum mechanics, the (classical) probability distribution over the answers to a particular experimental question---the one that we infer, conditional upon our asking that question---cannot be embedded into a global prior probability distribution over all of the answers to all of the questions we might want to ask. But although these individual probability distributions do not logically fit together as they do for a classical system, there are logical relations between them nonetheless, that are compactly described by the quantum state.

For instance, as a general rule (which admits exceptions) one cannot use a quantum system to compute a logical disjunction by computing the individual values of each disjunct. This is simply because in the general case both disjuncts will not be globally defined. However quantum mechanics' logical structure provides other ways to compute a logical disjunction, and these other ways (which are not available to a classical computer) are exploited by a quantum computer to compute certain problems more efficiently than a classical computer can compute them.

Another view is Armond Duwell's \citeyearpar[]{duwell2018, duwellCompPhys}, who in contrast, agrees with the proponent of the many-worlds explanation in identifying quantum parallelism as at least part of the source of the power of quantum computers. Duwell, however, resists the temptation to draw a metaphysical inference from parallelism to many computational worlds. For Duwell the source of the power of a quantum computer lies in the way that it can efficiently correlate multiple values of a function and use these correlations to efficiently extract global information about the function. The disagreement over whether a quantum computer performs more, or fewer, computations than a classical computer is one that Duwell views as arising from conflicting intuitions about how to appropriately describe the quantum systems that perform computational tasks.

In these and other candidate explanations of the power of quantum computers that one encounters in the philosophical literature on the topic, the fact that quantum-mechanical systems exhibit \emph{entanglement} (see Section \ref{s:quantum-states-and-operations}) invariably plays an important role. For Bub entanglement is absolutely central, as entanglement is a (direct) manifestation of the fact that the logical structures of quantum and classical mechanics differ. As for Duwell, he takes his quantum parallelism thesis to be completely compatible with the idea that entanglement plays a central role in quantum computing, even if his explanation emphasises the correlations between the values of the function being evaluated by a system rather than the logical structure of its underlying state space \citep[p. 101]{duwell2018}. The many worlds advocate, as well, views quantum entanglement to be indispensable in the analysis of the power of quantum computing \citep[p. 889]{hewittHorsman2009}, even if for the many-worlds advocate it does not, by itself, suffice as a philosophical explanation for it.

The debate over the interpretation of the phenomenon of quantum entanglement has historically been one of the central debates in the controversy over quantum theory's conceptual foundations. First emphasised by Albert Einstein, Boris Podolsky, and Nathan Rosen in their \citeyear{epr1935} criticism of the orthodox interpretation of quantum mechanics, Erwin Schr\"odinger called it ``\emph{the} characteristic trait of quantum mechanics, the one that enforces its entire departure from classical lines of thought'' \citep[p. 555, emphasis in the original]{schrodinger1935}. It was only with the work of John Bell \citeyearpar[]{bell1964, bell1966}, however, that its significance for our understanding of the break that quantum mechanics makes with classical physics was first made fully clear.

Consider Louise and Robert, two friends who live in the 15th and 11th arrondissements, on the left bank and the right bank of the Seine, respectively, in the city of Paris. Every morning, Louise and Robert each receive a letter in their respective mailboxes that consists of a single sheet of paper inscribed with either a large bass clef:
$$\fclef$$
or a large treble clef
$$\gclef.$$
After awhile, Louise determines that it is equally likely, on any given day, that she will receive a bass clef as it is that she will receive a treble clef. After awhile Robert determines the same. But when they compare their notes, they find that whenever Robert receives a bass clef, so does Louise. Similarly, every time he receives a treble clef, she does too. In other words, there is a one-in-two chance, on any given day, that they both receive a bass clef, and a one-in-two chance that they both receive a treble clef. No other combinations are ever observed. In other words, their outcomes are \emph{correlated}, such that the probability distribution over the possible combinations of outcomes is given by:\footnote{If the outcomes were completely uncorrelated, the probability distribution would be $$\frac{1}{4}\big[\fclef\big]_L\big[\fclef\big]_R \;+\; \frac{1}{4}\big[\fclef\big]_L\big[\gclef\big]_R \;+\; \frac{1}{4}\big[\gclef\big]_L\big[\fclef\big]_R \;+\; \frac{1}{4}\big[\gclef\big]_L\big[\gclef\big]_R.$$}
\begin{align}
\label{e:linda-robert}
\frac{1}{2}\big[\fclef\big]_L\big[\fclef\big]_R \;+\; \frac{1}{2}\big[\gclef\big]_L\big[\gclef\big]_R.
\end{align}

What explains this correlation? Well in this case it turns out that Louise and Robert both play in a jazz ensemble. The band's leader, Carsten, lives in the centre of the city (on the \^Ile Saint-Louis). Every afternoon he flips a fair coin. Depending on whether the coin lands heads or tails, he either writes a large bass clef or a large treble clef on a sheet of paper, photocopies it, and sends one copy each to Robert and Louise. If Robert receives a treble clef from Carsten, then he will know to take his tenor horn to the jazz club that night. Otherwise he will take his trombone. As for Linda, if she receives a treble clef, she will know to bring her soprano clarinet. Otherwise she will bring her bassoon.

The result of Carsten's coin flip is called the \emph{common cause} of Louise's and Robert's outcomes, and the story we tell about how Carsten's coin flip determines Louise's and Robert's outcomes is called a \emph{causal model} for the correlations that they see \citep[see][]{pearl2009}. If Louise and Robert do not know how Carsten is determining what to send them, then they will wonder about the \emph{hidden-variable} that explains their correlation. In this case the result of Carsten's coin flip is actually a \emph{local} hidden-variable, since the process by which the outcome of the coin flip determines what is written on the letters is confined to the small localised region in the vicinity of the desk in Carsten's central office. He flips the coin, takes note of the outcome (the local hidden-variable), and writes the corresponding symbol.

Instead of simply flipping a coin while sitting at his desk, we can imagine a more complicated, spatially distributed, process by which Carsten determines what to write. For instance Carsten might begin by flipping his coin, and then, corresponding to heads or tails he might telephone Tilde or Bjarne, who live in the northern and southern suburbs of the city respectively, and ask whichever of them he calls to roll a six-sided die at exactly seven-o'clock in the evening, and to afterwards call him back at either 7:13PM, if it is Bjarne, or 7:18PM, if it is Tilde, to tell him the outcome. Then, if the result of the die roll is one, three, or six, Carsten will write a treble clef on the sheet of paper before photocopying it and sending copies to Robert and Louise, while if it is two, four, or five, he will write a bass clef. All of these actions together constitute, just like the coin flip in our simpler scenario, a locally causal model to explain Louise's and Robert's correlation. Why do we call it local even though it is spatially distributed? Because the physical processes (the coin flips, die rolls, and telephone calls) by which Carsten determines what to write propagate locally in the \emph{physical sense}, i.e., at a finite speed that is less than the speed of light.

\citet[]{bell1964} showed, first, that there are certain probabilistic constraints---what are now called \emph{Bell inequalities}---on the statistics arising from measurements on \emph{any} locally correlated system. He then showed that these constraints are violated by statistics that are predicted to arise from certain experiments on a system composed of two qubits in the quantum-mechanical state:\footnote{This state is identical to the one given in Eq.\ \eqref{e:entanglement-comp} but we repeat it here for convenience.}
\begin{align}
  \label{e:entanglement-comp2}
  \frac{1}{\sqrt 2}\big(| 0 \rangle_A| 1 \rangle_B \;-\; | 1 \rangle_A| 0 \rangle_B\big),
\end{align}
where the $A$ and $B$ subsystems can be as far apart as one likes. The proof of this violation is known as \emph{Bell's theorem} and it, and its variants, have since been experimentally confirmed many times over \citep[]{genovese2016}.

The predicted violation only occurs for certain measurements. If we measure both the $A$ and $B$ qubits in the computational basis (see Eq.\ \eqref{eqn:general-form-state-vector} and Figure \ref{f:different-bases}), then the predicted statistics will actually be compatible with the constraints imposed by local hidden-variable theories, as Bell himself showed \citeyearpar[p. 16]{bell1964}. But as we rotate the measurement basis (see Figure \ref{f:different-bases}) that we use for $B$ away from the measurement basis that we use for $A$, quantum mechanics predicts that we will see a violation, that it will reach a maximum at a certain point, and then decrease again as the measurement basis for $B$ begins to line up again with the measurement basis for $A$.

Another name for the computational basis is the $Z$-basis. We call it the $Z$-basis because the two computational basis vectors $| 0 \rangle$ and $| 1 \rangle$ correspond to the two possible outcomes of a \emph{$Z$-basis measurement} on a qubit, where ``measuring a qubit in the $Z$-basis'' means sending it through a $Z$-gate (see Figure \ref{f:quantum-circuit-model}) and then recording its state \citep[\S 6.5]{3m2020}. Similarly, the basis vectors $| + \rangle$ and $| - \rangle$ are the two possible outcomes of an $X$-basis measurement on a qubit, and $| y^+ \rangle$ and $| y^- \rangle$ are the two possible outcomes of a $Y$-basis measurement. The $X$, $Y$, and $Z$ gates, together with the trivial $I$ gate that leaves a qubit's state unchanged, are known as the \emph{Pauli} gates.

I will have more to say about this family of gates later. For now I want to point out that, for a system of qubits in the state given by Eq.\ \eqref{e:entanglement-comp2}, as long as both the $A$ and $B$ qubit are measured in one of the Pauli bases, the predicted statistics arising from those measurements will not violate the constraints that Bell's inequality imposes on local hidden-variable theories. In other words, if all that we have access to are measurement devices that can measure a qubit in one of the Pauli bases, then there will be no way to disprove some local hidden-variable story a skeptic might cook up to explain the observed statistics. To experimentally disprove such a story, we will need to have measurement devices that can measure in other measurement bases besides these ones, the ones for which Bell showed that a violation of the Bell inequalities will occur.

Bell's theorem should convince us that no local hidden-variable theory can reproduce the correlations arising from such measurements (i.e., from measurements in bases other than $X$, $Y$, and $Z$), but this does not bar a skeptic from positing a \emph{nonlocal} hidden-variable theory to recover them instead. An example of such a theory is one in which the outcome of a measurement on $A$ depends on the measurement basis used to measure $B$. But since $A$ and $B$ could in principle be far apart in space, and since we require the measurements to be performed simultaneously, the causal influence from $B$ to $A$ in such a theory will have to be propagated faster than the speed of light. This is a hard pill to swallow, but one might be inclined to swallow it anyway in order to avoid the other alternatives.\footnote{For further discussion, see \citet{myrvoldGenoveseShimonySEP}.}

Coming back to our discussion of quantum computers, recall that when we discussed the many-worlds explanation of quantum computing, we noted that the computer's state as given on the right-hand side of Eq.\ \eqref{e:parallel} is entangled. The question arises, then, as to what role is played by entanglement more generally in enabling quantum computers to outperform their classical competitors. There are two ways to think of this question. We might ask, on the one hand, whether realising an entangled state is \emph{necessary} to enable a quantum speedup. On the other hand we might ask whether it is \emph{sufficient} to enable it.

There has been some debate surrounding the first question. One the one hand, \citet[]{jozsa2003} have proven that realising an entangled state is necessary to enable speedup if a quantum computer can be assumed to be in a so-called `pure state', i.e., in a state that represents a maximally specific description of the system from the point of view of the quantum-mechanical formalism. This is the case when we represent a system with a state vector as we have been doing up until now (and will continue to do). On the other hand it has been argued \citep[]{biham2004} that quantum computers that are in `mixed states'---i.e., states which describe the computer in a less than maximally specific way, either because we are ignorant of the actual state of the computer, or because the computer is coupled to its environment---are in some cases capable of achieving a quantum speedup over their classical competitors without ever being entangled. There is insufficient space to discuss this (largely technical) debate here, but in \citet[]{cuffaro2013b} I argue that the purported counter-examples to what I there call the ``necessity of entanglement thesis'' do not actually demonstrate what they purport to show, but instead clarify the necessary role that entanglement does play in quantum computation.

From the philosopher's point of view, the more interesting question in relation to the role of entanglement in quantum computing is the question regarding sufficiency, for as we will see, reflecting on this question sheds light on the tension between physical and computational ways of thinking that is at the heart of the science of quantum computing, the tension that is the primary source of the insight this new science brings into both physics and computation.

\section{The Gottesman-Knill theorem}
\label{s:gk-theorem}

The main reason for being skeptical of the idea that entanglement suffices to enable quantum speedup is the Gottesman-Knill theorem \citep[]{gottesman1999}. This theorem states that any quantum algorithm that employs (exclusively) some combination of the following operations (which together form what we will call the \emph{Gottesman-Knill set}) is efficiently simulable by a classical computer: (i) The Clifford group of gates (see Figure \ref{f:quantum-circuit-model}), i.e., the $X$, $Y$, $Z$, $R$, $H$, and ${\rm CNOT}$ gates; (ii) Clifford group gates conditioned on the values of classical bits (indicating, e.g., the results of previous measurements); (iii) state preparation of a qubit in the computational basis (as one typically does for each qubit prior to the beginning of a computation); (iv) measurements in one of the Pauli bases (as one does at the end of a computation) \citep[\S 10.5.4]{nielsenChuang2000}.

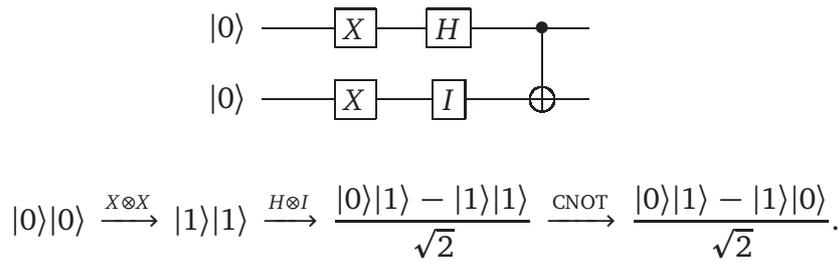
\begin{figure}
  \begin{align*}
    \Qcircuit @C=.1em @R=.91em @! {
    \lstick{| 0 \rangle} & \qw & \gate{X} & \qw & \gate{H} & \qw & \ctrl{1} & \qw \\
    \lstick{| 0 \rangle} & \qw & \gate{X} & \qw & \gate{I} & \qw & \targ    & \qw \\
    }
  \end{align*}
  \begin{align*}
    | 0 \rangle| 0 \rangle \; \xrightarrow{X \otimes X} \;
    | 1 \rangle| 1 \rangle \; \xrightarrow{H \otimes I} \;
    \frac{| 0 \rangle| 1 \rangle \, - \, | 1 \rangle| 1 \rangle}{\sqrt 2} \; \xrightarrow{{\rm CNOT}} \;
    \frac{| 0 \rangle| 1 \rangle \, - \, | 1 \rangle| 0 \rangle}{\sqrt 2}.
  \end{align*}
  \caption{A quantum circuit (above), written out explicitly in terms of the unitary transformations required to realise it (below), involving only operations from the Gottesman-Knill set. At the end of this sequence of operations the computer will be in an entangled state.}
  \label{f:gk-entanglement}
\end{figure}

The reason the theorem represents a challenge, to the view that entanglement suffices to enable a quantum computer to outperform a classical computer, is that by using only the operations given in (i)--(iv), which according to the theorem are all efficiently classically simulable, it is possible to generate an entangled state. Figure \ref{f:gk-entanglement} depicts a quantum circuit to realise an entangled state using only operations from the Gottesman-Knill set. It begins with the preparation of two qubits in the computational basis state $| 0 \rangle$, then subjects them each to an $X$-gate, as a result of which they will both be in the state $| 1 \rangle$. The first qubit is then run through a Hadamard gate which transforms it into the superposition state $\nicefrac{(| 0 \rangle \, - \, | 1 \rangle)}{\sqrt 2}$, and following that both qubits are run through a ${\rm CNOT}$ gate. The resulting state of the computer is an entangled state.

It would be wrong to conclude \citep[cf.][pp. 2029--30]{jozsa2003}, on the basis of this, that entanglement is not sufficient to enable a quantum computer to outperform a classical computer. Why? Well, let us reflect on what the Gottesman-Knill theorem is saying. The Gottesman-Knill theorem shows that there is a certain set of quantum operations that can be efficiently simulated on a classical computer. Let us then consider what we mean when we say that we have `efficiently classically simulated' something. We can say, to start with, that in a classical computer simulation (efficient or otherwise), whatever computer is doing the simulating will be such that it can be given a classical description. As we saw in Section \ref{s:classical-states}, a complete description of a system, in classical mechanics, is always factorisable into complete descriptions of all of its subsystems (see Figure \ref{f:switches}). Besides this, classically describable systems and processes are locally causal, just like the convoluted procedure that Carsten uses to determine what to stuff in the envelopes he sends to Louise and Robert. The classical language of classical mechanics \emph{constrains} all the descriptions of systems that can be given in that language to be this way. The behaviour of a classically describable system, in other words, is \emph{always} expressible in terms of local causes and effects, and correlations between effects manifested by such a system's subsystems can always be explained by appealing either to common or direct causes, whose influences propagate no faster than the speed of light. This is true no matter how large a system is conceived to be. For classical mechanics aims to be a universal language for describing \emph{any} physical system, be it a single particle or the entire physical universe. The local causes of an effect will not always be known to us, but classical mechanics tells us that they are there, and that we can find them in principle.

Let us return to Carsten, our band leader who lives on the \^Ile Saint-Louis in the centre of Paris. Carsten has lately become interested in quantum computers as a way of helping with composition,\footnote{For a review of some of the uses envisioned for quantum computers in music, see \citet[]{mirandaHelloMusic}.} and has acquired a shiny silver box, now sitting on his desk, which he claims is one. Quadeisha, a friend and fellow musician who has happened by, is skeptical, and asks him to show her how it works. Obligingly, Carsten turns on his computer and has it run what he tells her is a calculation whose result depends on the prior generation of an entangled state. Quadeisha, still not satisfied, asks Carsten to describe the algorithm by which the computer arrives at its result. Upon discovering (after looking through the owner's manual for the device) that the algorithm employs no operations outside of the Gottesman-Knill set, she refuses to believe that Carsten's computer is quantum.

Quadeisha is right to be skeptical, \emph{even if} the box on Carsten's desk (which, let us assume, he does not know how to open) \emph{actually is} a quantum computer.\footnote{This is actually the attitude (and for similar reasons) that many theorists take toward certain claims by private corporations to have built a working quantum computer (see, e.g., \citet{aaronson2013a}; \citet{shin2014}).} Moreover she need not have heard of the Gottesman-Knill theorem to be justified in her skepticism. She need only be familiar with the work of John Bell. For it can be shown that the combined effect of any sequence of Gottesman-Knill operations, for any subsystem of the system to which they are applied, is equivalent to a measurement in one of the Pauli bases, $X$, $Y$, $Z$, on a system whose state is given by one of the basis vectors of a Pauli basis \citep[][p. 115]{cuffaro2017}. Bell already showed us how to provide a locally causal model to reproduce the statistics of such measurements on a system in the state given in Eq.\ \eqref{e:entanglement-comp2}. Moreover his technique is straightforwardly extendable to other similarly entangled two-party systems \citep[][\S A.1]{cuffaro2017}, and further techniques have been devised for constructing efficient locally causal models to recover the statistics arising from Pauli-basis measurements on entangled systems composed of three or more parties as well (see \citeauthor{cuffaro2017}, \citeyear[][\S A.2]{cuffaro2017}, which summarises a technique first described in \citeauthor{tessier2004}, \citeyear[]{tessier2004}).\footnote{The $n$-party case, for $n \geq 3$, introduces subtleties which we will discuss in the next section.}

The upshot of the Gottesman-Knill theorem is that a certain number of quantum operations are efficiently classically simulable despite being capable of generating entangled states. And we have just seen that to say that some phenomenon is `efficiently classically simulable' is to say that it can be efficiently redescribed in a locally causal way. Since any sequence of Gottesman-Knill operations is equivalent to a measurement in one of the Pauli bases, $X$, $Y$, $Z$, on a system whose state is given by one of the basis vectors of a Pauli basis, the Gottesman-Knill theorem is essentially telling us that there are some statistics associated with measurements on systems in entangled states that admit of a locally causal description.

We do not really need the Gottesman-Knill theorem to tell us this, for Bell's and related inequalities amount to general constraints that any locally causal model of combined measurement statistics must satisfy, and we already know that in some cases, i.e., for measurements in one of the Pauli bases, the predictions of a locally causal model are compatible with quantum-mechanical predictions. It is therefore misleading to conclude, on the basis of the Gottesman-Knill theorem, that entanglement is not sufficient to enable quantum computational speedup. What the Gottesman-Knill theorem shows us is that one also needs to \emph{use} this entanglement to its full potential, in particular by not restricting a quantum computer to use only a small proportion of the operations of which it is capable (i.e., by not restricting it to the Gottesman-Knill set).

In the context of more general discussions of quantum mechanics, there are those who claim that entanglement is \emph{the} one and only distinguishing feature of the theory \citep{schrodinger1935}. This is a controversial claim, but it is not proved false by the fact that \citet[]{bell1964} showed that some operations on some entangled states can be redescribed in a locally causal way. Likewise, pointing out essentially the same thing in the context of quantum computation is not an objection to the claim that quantum entanglement constitutes a sufficient physical resource for realising speedup over classical computation. Indeed, the Gottesman-Knill theorem serves to highlight the role that is actually played by entanglement in a quantum computer and to clarify why and in what sense it is sufficient to preclude the computer's evolution from plausibly being classically simulated \citep[for more on this, see][]{cuffaro2017}.

\section{Computer simulations and locally causal models of quantum phenomena}
\label{s:computer-simulations}

Above we mentioned that techniques have been developed for providing classical computer simulations, i.e., locally causal descriptions, to recover the statistics arising from Pauli-basis measurements on entangled systems whose subsystems are in superpositions of eigenstates of Pauli observables, for any number of parties. The classical computer simulations of three or more party systems are especially interesting in that they necessarily involve a small amount of communication. For three parties, $A$, $B$, and $C$, one bit must be communicated between $B$ and $A$ in order to make the protocol work. For $n$ parties, $n-2$ bits are required, i.e., the number of bits required scales linearly with $n$ \citep[]{tessier2004}. Note that this counts as being `easy' in the complexity-theoretic sense we discussed in Section \ref{s:classical-computers}.

From a certain point of view, however, it may seem quite odd to call a classical computer simulation of quantum correlations, that involves a certain amount of communication between parties, a locally causal model of those correlations. Recall that, earlier, when we introduced Bell's theorem, we claimed that communication between distant parties in the context of a Bell experiment constitutes a \emph{nonlocal} influence, i.e., one that would have to be communicated at a speed exceeding that of light.

We can start to resolve the tension here if we consider, first, just what a locally causal redescription of quantum measurement statistics is required to do. If a pair of qubits, $A$ and $B$, are prepared in an entangled state and then spatially separated, quantum mechanics tells us that when, for instance, a $Z$-basis measurement is performed on $B$, $A$'s state changes `instantaneously', so that a $Z$-basis measurement on $A$ will yield an outcome that is correlated with the result of the experiment on $B$ in a particular way. Our challenge, however, is not to reproduce these apparent dynamics. Our challenge is to come up with an alternative description of the statistics that we \emph{actually observe} in our experiments.

In order to assert that one has actually observed the results of a \emph{combined} experiment on two or more spatially separated subsystems, it must be assumed that the results associated with the individual experiments on those subsystems have actually been combined in some way. That is, we must somehow gather together the results registered locally at the sites of the experiments on $A$ and $B$. If Alice (a bassist) is assigned to measure $A$, for instance, and Bob (who plays the piano) to measure $B$, then once they record their individual results they need to communicate them, either to one another, or perhaps to Carsten who is sitting in his office on the \^Ile Saint-Louis, where he is in the process of showing Quadeisha how his quantum computer is but one part of a network of spatially distributed quantum computers that are together performing a computation that depends on the prior generation of a system in an entangled state. While Alice and Bob are making their way toward Carsten and Quadeisha with their results, however, there is time for Alice and Bob to, surreptitiously, exchange further information with one another by local means (Bob might call Alice on his cell phone, for instance, as they are making their way toward Carsten's office) in order to coordinate the outcomes of their individual measurements and `correct' them if necessary, unbeknownst to Quadeisha. And since this further information is propagated locally, i.e., at a speed no greater than that of light, it can be considered as \emph{part of the common cause} of the actually observed (combined) measurement event, and thus as part of a classical, locally causal, description of that measurement event \citep[see][\S 4, fig. 1]{cuffaro2017}.

Now, given a particular combined calculational outcome at a time $t$, one can try to explain it in one of two ways: (i) Carsten's way, i.e., as the result of a \emph{bona fide} quantum-mechanical process manifesting nonlocal behaviour, or (ii) Quadeisha's more skeptical way, i.e., as the result of a locally causal mechanism whose different computational components are communicating with one another at less than the speed of light. Without further evidence---\emph{if Carsten cannot or will not open his silver box}---it is perfectly legitimate to side with Quadeisha here, and be a skeptic with respect to Carsten's quantum-mechanical description of the box's observed behaviour. Quadeisha's alternative account of the observed correlations is a perfectly good one in such a scenario, for her proposed classical alternative can solve the problem just as easily, in the complexity-theoretic sense, as Carsten's purported quantum computer.

Bell's and related inequalities specify \emph{constraints} on what a locally causal description that accounts for a combined probability distribution must be like, but in and of themselves they are essentially little more than formal statements. Bell's \citeyearpar{bell1964} original inequality, for instance, is essentially just a theorem of probability (\citealt[]{bell1981}, \citealt[\S 3.4]{3m2020}, \citealt{pitowsky1989}, \citealt{pitowsky1994}). This is no fault. But if we are to make a meaningful distinction between what is and is not ruled out by them, then we must consider the context within which they are being applied. Normally this does not need to be made explicit, for normally the context is what I have elsewhere called the \emph{theoretical} context \citep[p. 107]{cuffaro2017}, wherein we are considering alternative theories of the natural world. In this context, Bell's inequalities help us to answer the question of whether there may be some deeper hidden-variable theory of the natural world underlying quantum mechanics, and what such a theory needs to be like.

They do not answer that question all by themselves, however. Any alternative hidden-variable theory of the natural world must do more than merely satisfy the constraints imposed by the Bell inequalities. It must also be \emph{plausible}. Thus, besides reproducing the well-confirmed statistical predictions of quantum mechanics, any deeper candidate theory should also, for instance, be consistent with our other well-confirmed theories of physics such as special and general relativity (and if not, then a compelling reason will need to be given for why we should not worry about such contradictions). It is on the basis of such plausibility constraints, and not on the basis of the Bell inequalities, that we rule out many of the so-called `loopholes' to Bell's theorem. Quadeisha's alternative account of the observed correlations in Carsten's office counts as a particularly implausible example of such a loophole in the theoretical context.\footnote{Quadeisha's loophole is actually conceptually similar to the `collapse locality' loophole at the heart of Adrian Kent's \emph{causal quantum theory} \citep[]{kent2005}. For discussion see \citet[pp. 104--106]{cuffaro2017}. For a more general discussion of the methodology of no-go theorems, see \citet[]{dardashti2021}.}

But the theoretical context is not the only context that the Bell inequalities are relevant to. Nor is it the context appropriate to a discussion of the respective capabilities of quantum and classical computers. The context that is appropriate here is what I have elsewhere called the \emph{practical} context \citep[p. 107]{cuffaro2017}. In the practical context what concerns us is not alternative theories of the natural world. What concerns us is what \emph{we} are capable of \emph{building} with the aim of reproducing the statistical predictions of quantum mechanics. The plausibility constraints that restrict us in the practical context are not the same as the ones that restrict us in the theoretical context. But there are plausibility constraints in the practical context nevertheless.

In particular, in the practical context, we should rule out, as implausible, alternative locally causal descriptions of systems that would be \emph{too hard} (in the complexity-theoretic sense) for us to build. And we should rule in alternative locally causal descriptions of systems that can be built \emph{easily} (again, in the complexity-theoretic sense). The complexity involved in the specification of such a system, in other words, needs to be \emph{tractable}. If, for instance, Quadeisha's basis for being skeptical of Carsten's claim to own a quantum computer is that she can specify a locally causal model for his computer's calculation that requires a number of additional resources that scales exponentially, then Carsten will be justified in dismissing her skepticism, for he can with justice say that it is wildly implausible that anyone can have built a machine according to Beatrice's specifications, because simulating the quantum calculation would amount, for such a machine, to an intractable task \citep[\S 5]{cuffaro2017}.\footnote{Note that I am taking `tractable' here in a relative sense. That is, the resources required by a classical computer to reproduce a particular effect should differ tractably from those required by a quantum computer. Or in other words: it should \emph{not be essentially harder} for the classical system to produce the same effect as the quantum system.}

\section{Computational perspectives on physics}
\label{s:pccp1}

In the previous section we discussed classical computer simulations of quantum-mechanical phenomena, and we made the point that such simulations can be thought of as locally causal alternative descriptions of those phenomena. This is not the way that they are normally framed in the literature on quantum information and computation. The way that it is normally put is that such simulations quantify the extent to which quantum mechanics differs from classical mechanics, by enumerating the number of resources required to account for quantum-mechanical phenomena using classical means (see, for instance \citealt{tessier2004}, \citealt{tessier2005}, \citealt{toner2003}, \citealt{brassard1999}, \citealt{rosset2013}).

Framing their significance in this way is certainly not incorrect, and it can be very theoretically useful.\footnote{The general subject of classical simulations of quantum systems is an important and burgeoning area of modern physics (see, for example, \citealt{lee2002}; \citealt{borges2010}).} But from the philosopher's perspective there is value in, so to speak, calling a spade a spade. A description of a plausible classical computational model for efficiently simulating some of the operations that can be carried out by a quantum computer is the same kind of thing as a plausible alternative local hidden-variable account of some particular class of quantum phenomena, in the simple sense that in both cases these constitute plausible locally causal models of quantum phenomena. The difference between them lies in the way that one interprets the word `plausible' in each case. In the practical context we assume that a system has been built by a rational agent to achieve a particular purpose. We do not presuppose this in the theoretical context.\footnote{This statement is not meant to express any sort of theological opinion. It is merely a statement about how science operates, at least in this century.} Clearly, certain assumptions that would be regarded as plausible in the former context will not be in the latter and it is important to be on guard against conflating them. As we saw in the previous section, practical investigators attempting to isolate and/or quantify the computational resources provided by physical systems may be in danger of conceptual confusion if they are not cognisant of the differences between the practical and theoretical contexts. More positively, it must not be assumed by practical investigators that every `no-go result', formulated in the theoretical context, will constrain what they can accomplish in the practical context \citep[]{cuffaro2018a}. More generally it is important to think seriously about what one means by a `plausible locally causal model' regardless of the context of investigation.

Calling a spade a spade gives us, moreover, insight into how the practical context can illuminate the investigations of the traditional philosopher of physics in the context of quantum mechanics. The silly story we told in the previous section about how Carsten, Alice, and Bob manage to reproduce, by communicating just a few classical bits, all of the correlational phenomena associated with Pauli-basis measurements on an $n$-party quantum system in an entangled state, besides representing just another silly classical story, emphasises perhaps the most fundamental difference that we can point to between classical and quantum mechanics. Classical and quantum mechanics are, fundamentally, alternative universal languages for describing physical systems \citep[\S 6.3]{3m2020}. And by investigating the computational power inherent in those languages we gain insight into the respective logical structures of classical and quantum theory, into the logical structure of quantum mechanics that enables the efficient representation of correlational phenomena by quantum-mechanical systems, and the logical structure of classical mechanics which precludes this \citep[\S 4.2]{3m2020}.

\section{Concluding remarks}
\label{s:conclusion}

Reflecting on the lessons of quantum mechanics, Niels Bohr wrote that:

\begin{quote}
In representing a generalization of classical mechanics suited to allow for the existence of the quantum of action, Quantum mechanics offers a frame \emph{sufficiently wide} to account for empirical regularities \emph{which cannot} be comprised in the classical way of description \citep[p.\ 316, my emphasis]{bohr1948}.
\end{quote}

This is a lesson with which the artist will be familiar.\footnote{A similar point is made in \citet[sec. 6.3, note 22]{3m2020}.} From time to time, in literature, music, and in other forms of art, new methods of writing and composing emerge that allow one to express the subtleties and nuances of experience more easily than before. E.\ M.\ \citet[p.\ 28]{Forster_1942} once said, about Virginia Woolf, that ``She pushed the light of the English language a little further against the darkness.'' In music, the (re-)introduction, into the Western musical tradition, of modality and chromaticism through Jazz, Blues, and other forms of popular music, and also through art music \citep[see][]{vincent1951}, has enabled modern composers in the Western tradition to explore musical landscapes and express aspects of our emotional and intellectual experience that composers limited to the major and minor scales, so dominant during Western music's classical period, cannot easily capture. In his musings about quantum mechanics, Bohr himself was wont to appeal to an analogy with poetry. Werner Heisenberg recalls, in his memoirs, a conversation he had with Bohr in which the latter stated that:

\begin{quote}
We must be clear that, when it comes to atoms, language can be used only as poetry. The poet, too, is not nearly so concerned with describing facts as with creating images and establishing mental connections'' \citep[p. 41]{heisenberg1971}.  
\end{quote}

These and like analogies were not lost on actually practising poets and other artists of the same period \citep[see][]{mairhofer2021}. The subject of quantum mechanics has even been found to lend itself very naturally to the format of the comic book (see \citeauthor{bub2018}, \citeyear{bub2018}, and the review by \citeauthor{cuffaroDoyle2021}, \citeyear[especially \S 4]{cuffaroDoyle2021}). As for the science of quantum computing: The science of quantum computing takes advantage, as we have seen throughout the course of this chapter, of the increased expressive power of the quantum-mechanical language. It shows us that using the new language of quantum mechanics, we can interact with what we take to be quantum-mechanical systems in ways that cannot be easily replicated classically.

In the course of this chapter we have reflected on the fundamentals of classical and quantum physics and computation, on the resources used and on the explanation of the power of quantum computers, and finally on the broader insights that can be gleaned from the science of quantum computing both for physics and for computation. The philosophical issues addressed in this chapter are not the only issues brought up by the science of quantum computing. But I hope to have convinced the reader, with this sample of some of the more central ones, of quantum computing's potential for illuminating the world that we are all participants in.

\bibliographystyle{apa-good}
\bibliography{qcSpringer}{}

\begin{thebibliography}{124}
\expandafter\ifx\csname natexlab\endcsname\relax\def\natexlab#1{#1}\fi
\expandafter\ifx\csname url\endcsname\relax
  \def\url#1{{\tt #1}}\fi
\expandafter\ifx\csname urlprefix\endcsname\relax\def\urlprefix{URL }\fi

\bibitem[{Aaronson(2013{\natexlab{a}})}]{aaronson2013a}
Aaronson, S. (2013{\natexlab{a}}).
\newblock {D-Wave}: Truth finally starts to emerge.
\newblock Posted: June 05, 2013. Retrieved: August 11, 2014.
\newline\urlprefix\url{www.scottaaronson.com/blog/?p=1400}

\bibitem[{Aaronson(2013{\natexlab{b}})}]{aaronson2013}
Aaronson, S. (2013{\natexlab{b}}).
\newblock {\em Quantum Computing Since {D}emocritus\/}.
\newblock New York: Cambridge University Press.

\bibitem[{Aaronson(2013{\natexlab{c}})}]{aaronson2013b}
Aaronson, S. (2013{\natexlab{c}}).
\newblock Why philosophers should care about computational complexity.
\newblock In B.~J. Copeland, C.~J. Posy, \& O.~Shagrir (Eds.) {\em
  Computability: {T}uring, {G}\"odel, {C}hurch, and Beyond\/}, (pp. 261--327).
  Cambridge, MA: MIT Press.

\bibitem[{Aaronson(2016)}]{aaronson2012}
Aaronson, S. (2016).
\newblock Complexity zoo.
\newblock \url{complexityzoo.uwaterloo.ca/}\\\url{Complexity\_Zoo}.

\bibitem[{Adlam(2014)}]{adlam2014}
Adlam, E. (2014).
\newblock The problem of confirmation in the {E}verett interpretation.
\newblock {\em Studies in History and Philosophy of Modern Physics\/}, {\em
  47\/}, 21--32.

\bibitem[{Agrawal et~al.(2004)Agrawal, Kayal, \& Saxena}]{agrawal2004}
Agrawal, M., Kayal, N., \& Saxena, N. (2004).
\newblock {PRIMES} is in {P}.
\newblock {\em Annals of Mathematics\/}, {\em 160\/}, 781--793.

\bibitem[{Aharonov et~al.(2007)Aharonov, van Dam, Kempe, Landau, Lloyd, \&
  Regev}]{aharanov2007}
Aharonov, D., van Dam, W., Kempe, J., Landau, Z., Lloyd, S., \& Regev, O.
  (2007).
\newblock Adiabatic quantum computation is equivalent to standard quantum
  computation.
\newblock {\em {SIAM} Journal on Computing\/}, {\em 37\/}, 166--194.

\bibitem[{Albert \& Loewer(1988)}]{albert1988}
Albert, D., \& Loewer, B. (1988).
\newblock Interpreting the many worlds interpretation.
\newblock {\em Synthese\/}, {\em 77\/}, 195--213.

\bibitem[{Arora \& Barak(2009)}]{arora2009}
Arora, S., \& Barak, B. (2009).
\newblock {\em Computational Complexity: A Modern Approach\/}.
\newblock Cambridge: Cambridge University Press.

\bibitem[{Bell(2004 {[1964]})}]{bell1964}
Bell, J.~S. (2004 {[1964]}).
\newblock On the {Einstein-Podolsky-Rosen} paradox.
\newblock In {\em Speakable and Unspeakable in Quantum Mechanics\/}, (pp.
  14--21). Cambridge: Cambridge University Press.

\bibitem[{Bell(2004 {[1966]})}]{bell1966}
Bell, J.~S. (2004 {[1966]}).
\newblock On the problem of hidden variables in quantum mechanics.
\newblock In {\em Speakable and Unspeakable in Quantum Mechanics\/}, (pp.
  1--13). Cambridge: Cambridge University Press.

\bibitem[{Bell(2004 {[1981]})}]{bell1981}
Bell, J.~S. (2004 {[1981]}).
\newblock Bertlmann's socks and the nature of reality.
\newblock In {\em Speakable and Unspeakable in Quantum Mechanics\/}, (pp.
  139--158). Cambridge: Cambridge University Press.

\bibitem[{Bell \& Gao(2016)}]{bellGao2016}
Bell, M., \& Gao, S. (Eds.)  (2016).
\newblock {\em Quantum Nonlocality and Reality\/}.
\newblock Cambridge: Cambridge University Press.

\bibitem[{Bennett et~al.(1997)Bennett, Bernstein, Brassard, \&
  Vazirani}]{bennett1997}
Bennett, C.~H., Bernstein, E., Brassard, G., \& Vazirani, U. (1997).
\newblock Strengths and weaknesses of quantum computing.
\newblock {\em SIAM Journal on Computing\/}, {\em 26\/}, 1510--1523.

\bibitem[{Bernstein \& Vazirani(1997)}]{bernstein1997}
Bernstein, E., \& Vazirani, U. (1997).
\newblock Quantum complexity theory.
\newblock {\em SIAM Journal on Computing\/}, {\em 26\/}, 1411--1473.

\bibitem[{Biham et~al.(2004)Biham, Brassard, Kenigsberg, \& Mor}]{biham2004}
Biham, E., Brassard, G., Kenigsberg, D., \& Mor, T. (2004).
\newblock Quantum computing without entanglement.
\newblock {\em Theoretical Computer Science\/}, {\em 320\/}, 15--33.

\bibitem[{Bohr(1948)}]{bohr1948}
Bohr, N. (1948).
\newblock On the notions of causality and complementarity.
\newblock {\em Dialectica\/}, {\em 2\/}, 312--319.

\bibitem[{Boole(1847)}]{boole1847}
Boole, G. (1847).
\newblock {\em The Mathematical Analysis of Logic\/}.
\newblock New York: Philosophical Library.
\newblock Reprinted: Bristol: Thoemmes Press, 1998.

\bibitem[{Borges et~al.(2010)Borges, Hor-Meyll, Huguenin, \&
  Khoury}]{borges2010}
Borges, C. V.~S., Hor-Meyll, M., Huguenin, J. A.~O., \& Khoury, A.~Z. (2010).
\newblock {B}ell-like inequality for the spin-orbit separability of a laser
  beam.
\newblock {\em Physical Review A\/}, {\em 82\/}, 033833.

\bibitem[{Brassard et~al.(1999)Brassard, Cleve, \& Tapp}]{brassard1999}
Brassard, G., Cleve, R., \& Tapp, A. (1999).
\newblock Cost of exactly simulating quantum entanglement with classical
  communication.
\newblock {\em Physical Review Letters\/}, {\em 83\/}, 1874--1877.

\bibitem[{Briegel et~al.(2009)Briegel, Browne, D\"ur, Raussendorf, \& den
  Nest}]{briegel2009}
Briegel, H.~J., Browne, D.~E., D\"ur, W., Raussendorf, R., \& den Nest, M.~V.
  (2009).
\newblock Measurement-based quantum computation.
\newblock {\em Nature Physics\/}, {\em 5\/}, 19--26.

\bibitem[{Bub(2006)}]{bub2006}
Bub, J. (2006).
\newblock Quantum computation from a quantum logical perspective.
\newblock {arXiv:quant-ph/0605243v2}.

\bibitem[{Bub(2008)}]{bub2008}
Bub, J. (2008).
\newblock Quantum computation and pseudotelepathic games.
\newblock {\em Philosophy of Science\/}, {\em 75\/}, 458--472.

\bibitem[{Bub(2010)}]{bub2010}
Bub, J. (2010).
\newblock Quantum computation: Where does the speed-up come from?
\newblock In A.~Bokulich, \& G.~Jaeger (Eds.) {\em Philosophy of Quantum
  Information and Entanglement\/}, (pp. 231--246). Cambridge: Cambridge
  University Press.

\bibitem[{Bub(2016)}]{bub2016}
Bub, J. (2016).
\newblock {\em Bananaworld, Quantum Mechanics for Primates\/}.
\newblock Oxford: Oxford University Press.

\bibitem[{Bub \& Pitowsky(2010)}]{bub-pitowsky2010}
Bub, J., \& Pitowsky, I. (2010).
\newblock Two dogmas about quantum mechanics.
\newblock In S.~Saunders, J.~Barrett, A.~Kent, \& D.~Wallace (Eds.) {\em Many
  Worlds? {E}verett, Quantum Theory, and Reality\/}, (pp. 433--459). Oxford:
  Oxford University Press.

\bibitem[{Bub \& Bub(2018)}]{bub2018}
Bub, T., \& Bub, J. (2018).
\newblock {\em Totally Random: Why Nobody Understands Quantum Mechanics\/}.
\newblock Princeton: Princeton University Press.

\bibitem[{Church(1936)}]{church1936}
Church, A. (1936).
\newblock An unsolvable problem of elementary number theory.
\newblock {\em American Journal of Mathematics\/}, {\em 58\/}, 345--363.

\bibitem[{Cobham(1965)}]{cobham1965}
Cobham, A. (1965).
\newblock The intrinsic computational difficulty of functions.
\newblock In Y.~Bar-Hillel (Ed.) {\em Logic, Methodology and Philosophy of
  Science: Proceedings of the 1964 International Congress\/}, (pp. 24--30).
  Amsterdam: North-Holland.

\bibitem[{Cook(2012)}]{cook2012}
Cook, W.~J. (2012).
\newblock {\em In Pursuit of the Traveling Salesman: Mathematics at the Limits
  of Computation\/}.
\newblock Princeton: Princeton University Press.

\bibitem[{Copeland(2017)}]{copeland2017}
Copeland, J.~B. (2017).
\newblock The modern history of computing.
\newblock In E.~N. Zalta (Ed.) {\em The {S}tanford Encyclopedia of
  Philosophy\/}. Metaphysics Research Lab, {S}tanford University, winter 2017
  ed.

\bibitem[{Copeland(2020)}]{copeland2020}
Copeland, J.~B. (2020).
\newblock The {C}hurch-{T}uring thesis.
\newblock In E.~N. Zalta (Ed.) {\em The {S}tanford Encyclopedia of
  Philosophy\/}. Metaphysics Research Lab, {S}tanford University, summer 2020
  ed.

\bibitem[{Cuffaro(2012)}]{cuffaro2012}
Cuffaro, M.~E. (2012).
\newblock Many worlds, the cluster-state quantum computer, and the problem of
  the preferred basis.
\newblock {\em Studies in History and Philosophy of Modern Physics\/}, {\em
  43\/}, 35--42.

\bibitem[{Cuffaro(2013)}]{cuffaro2013b}
Cuffaro, M.~E. (2013).
\newblock On the necessity of entanglement for the explanation of quantum
  speedup.
\newblock {arXiv:1112.1347v5}.

\bibitem[{Cuffaro(2017)}]{cuffaro2017}
Cuffaro, M.~E. (2017).
\newblock On the significance of the {G}ottesman-{K}nill theorem.
\newblock {\em The British Journal for the Philosophy of Science\/}, {\em
  68\/}, 91--121.

\bibitem[{Cuffaro(2018{\natexlab{a}})}]{cuffaro2018a}
Cuffaro, M.~E. (2018{\natexlab{a}}).
\newblock Reconsidering no-go-theorems from a practical perspective.
\newblock {\em The British Journal for the Philosophy of Science\/}, {\em
  69\/}, 633--655.

\bibitem[{Cuffaro(2018{\natexlab{b}})}]{cuffaro2018b}
Cuffaro, M.~E. (2018{\natexlab{b}}).
\newblock Universality, invariance, and the foundations of computational
  complexity in the light of the quantum computer.
\newblock In S.~O. Hansson (Ed.) {\em Technology and Mathematics: Philosophical
  and Historical Investigations\/}. Cham: Springer.

\bibitem[{Cuffaro(2020)}]{cuffaroInfCausality}
Cuffaro, M.~E. (2020).
\newblock Information causality, the {T}sirelson bound, and the `being-thus' of
  things.
\newblock {\em Studies in History and Philosophy of Modern Physics\/}, {\em
  72\/}, 266--277.

\bibitem[{Cuffaro \& Doyle(2021)}]{cuffaroDoyle2021}
Cuffaro, M.~E., \& Doyle, E.~P. (2021).
\newblock Essay review of {T}anya and {J}effrey {B}ub's \emph{Totally Random:
  Why Nobody Understands Quantum Mechanics: A Serious Comic on Entanglement}.
\newblock {\em Foundations of Physics\/}, {\em 51\/}, 28:1--28:16.

\bibitem[{Cuffaro \& Hartmann(2021)}]{cuffaroHartmannOpenSystems}
Cuffaro, M.~E., \& Hartmann, S. (2021).
\newblock The open systems view.
\newblock In preparation.

\bibitem[{Curiel(2014)}]{curiel2014}
Curiel, E. (2014).
\newblock Classical mechanics is {L}agrangian; it is not {H}amiltonian.
\newblock {\em The British Journal for Philosophy of Science\/}, {\em 65\/},
  269--321.

\bibitem[{Dardashti(2021)}]{dardashti2021}
Dardashti, R. (2021).
\newblock No-go theorems: What are they good for?
\newblock {\em Studies in History and Philosophy of Science\/}, {\em 86\/},
  47--55.

\bibitem[{Davis(2000)}]{davis2000}
Davis, M. (2000).
\newblock {\em The Universal Computer: The Road from Leibniz to Turing\/}.
\newblock New York: W. W. Norton and Company.

\bibitem[{Dawid \& Th\'ebault(2015)}]{dawidThebault2015}
Dawid, R., \& Th\'ebault, K. P.~Y. (2015).
\newblock Many worlds: Decoherent or incoherent?
\newblock {\em Synthese\/}, {\em 192\/}, 1559--1580.

\bibitem[{Dawson~Jr.(2007)}]{dawson2007}
Dawson~Jr., J.~W. (2007).
\newblock Classical logic's coming of age.
\newblock In D.~Jacquette (Ed.) {\em Philosophy of Logic\/}, (pp. 497--522).
  Amsterdam: Elsevier.

\bibitem[{Dean(2016)}]{dean2016b}
Dean, W. (2016).
\newblock Squeezing feasibility.
\newblock In A.~Beckmann, L.~Bienvenu, \& N.~Jonoska (Eds.) {\em Pursuit of the
  Universal: Proceedings of the 12th Conference on Computability in Europe\/},
  (pp. 78--88). Cham: Springer International Publishing.

\bibitem[{Deutsch(1985)}]{deutsch1985}
Deutsch, D. (1985).
\newblock Quantum theory, the {C}hurch-{T}uring principle and the universal
  quantum computer.
\newblock {\em Proceedings of the Royal Society of London. Series A.
  Mathematical and Physical Sciences\/}, {\em 400\/}, 97--117.

\bibitem[{Deutsch(1989)}]{deutsch1989}
Deutsch, D. (1989).
\newblock Quantum computational networks.
\newblock {\em Proceedings of the Royal Society of London. Series A.
  Mathematical and Physical Sciences\/}, {\em 425\/}, 73--90.

\bibitem[{Deutsch(1997)}]{deutsch1997}
Deutsch, D. (1997).
\newblock {\em The Fabric of Reality\/}.
\newblock New York: Penguin.

\bibitem[{Deutsch(2010)}]{deutsch2010}
Deutsch, D. (2010).
\newblock Apart from universes.
\newblock In S.~Saunders, J.~Barrett, A.~Kent, \& D.~Wallace (Eds.) {\em Many
  Worlds? Everett, Quantum Theory, and Reality\/}, (pp. 542--552). Oxford:
  Oxford University Press.

\bibitem[{DeWitt \& Graham(1973)}]{dewitt1973}
DeWitt, B., \& Graham, N. (1973).
\newblock {\em The Many-Worlds Interpretation of Quantum Mechanics\/}.
\newblock Princeton: Princeton University Press.

\bibitem[{{DeWitt}(1973 [1971])}]{dewitt1971}
{DeWitt}, B.~S. (1973 [1971]).
\newblock The many-universes interpretation of quantum mechanics.
\newblock In \citet[pp. 167--218]{dewitt1973}.

\bibitem[{Duwell(2007)}]{duwell2007}
Duwell, A. (2007).
\newblock The many-worlds interpretation and quantum computation.
\newblock {\em Philosophy of Science\/}, {\em 74\/}, 1007--1018.

\bibitem[{Duwell(2018)}]{duwell2018}
Duwell, A. (2018).
\newblock How to make orthogonal positions parallel: Revisiting the quantum
  parallelism thesis.
\newblock In M.~E. Cuffaro, \& S.~C. Fletcher (Eds.) {\em Physical Perspectives
  on Computation, Computational Perspectives on Physics\/}, (pp. 83--102).
  Cambridge: Cambridge University Press.

\bibitem[{Duwell(2021)}]{duwellCompPhys}
Duwell, A. (2021).
\newblock {\em Computation and Physics\/}.
\newblock Cambridge: Cambridge University Press.
\newblock Forthcoming.

\bibitem[{Edmonds(1965)}]{edmonds1965}
Edmonds, J. (1965).
\newblock Paths, trees, and flowers.
\newblock {\em Canadian Journal of Mathematics\/}, {\em 17\/}, 449--467.

\bibitem[{Einstein et~al.(1935)Einstein, Podolsky, \& Rosen}]{epr1935}
Einstein, A., Podolsky, B., \& Rosen, N. (1935).
\newblock Can quantum-mechanical description of physical reality be considered
  complete?
\newblock {\em Physical Review\/}, {\em 47\/}, 777--780.

\bibitem[{{Everett III}(1956)}]{everett1956}
{Everett III}, H. (1956).
\newblock The theory of the universal wave function.
\newblock In \citet[pp. 3--140]{dewitt1973}.

\bibitem[{Farhi et~al.(2000)Farhi, Goldstone, Gutmann, \& Sipser}]{farhi2000}
Farhi, E., Goldstone, J., Gutmann, S., \& Sipser, M. (2000).
\newblock Quantum computation by adiabatic evolution.
\newblock Tech. Rep. MIT-CTP-2936, MIT.
\newblock {arXiv}:quant-ph/0001106.

\bibitem[{Fletcher(2018)}]{fletcher2018}
Fletcher, S.~C. (2018).
\newblock Computers in abstraction / representation theory.
\newblock {\em Minds \& Machines\/}, {\em 28\/}, 445--463.

\bibitem[{Forster(1942)}]{Forster_1942}
Forster, E.~M. (1942).
\newblock {\em {V}irginia {W}oolf. The {R}ede Lecture 1941\/}.
\newblock Cambridge: Cambridge University Press.

\bibitem[{Genovese(2016)}]{genovese2016}
Genovese, M. (2016).
\newblock Experimental tests of {B}ell's inequalities.
\newblock In \citet[pp. 124--140]{bellGao2016}.

\bibitem[{G\"odel(1956)}]{godel1956}
G\"odel, K. (1956).
\newblock Private letter to {J}ohn {v}on {N}eumann, 20 {M}arch 1956.
\newblock Translated by Wensinger in: \cite{sipser1992}.

\bibitem[{Goldreich(2008)}]{goldreich2008}
Goldreich, O. (2008).
\newblock {\em Computational Complexity: A Conceptual Perspective\/}.
\newblock Cambridge: Cambridge University Press.

\bibitem[{Gottesman(1999)}]{gottesman1999}
Gottesman, D. (1999).
\newblock The {H}eisenberg representation of quantum computers.
\newblock In S.~P. Corney, R.~Delbourgo, \& P.~D. Jarvis (Eds.) {\em Group22:
  Proceedings of the {XXII} International Colloquium on Group Theoretical
  Methods in Physics\/}, (pp. 32--43). Cambridge, MA: International Press.
\newblock Longer version available at: {arXiv:quant-ph/9807006v1}.

\bibitem[{Greaves \& Myrvold(2010)}]{greavesMyrvold2010}
Greaves, H., \& Myrvold, W. (2010).
\newblock Everett and evidence.
\newblock In S.~Saunders, J.~Barrett, A.~Kent, \& D.~Wallace (Eds.) {\em Many
  Worlds? {E}verett, Quantum Theory, and Reality\/}, (pp. 181--205). Oxford:
  Oxford University Press.

\bibitem[{Grover(1996)}]{grover1996}
Grover, L.~K. (1996).
\newblock A fast quantum mechanical algorithm for database search.
\newblock In {\em Proceedings of the Twenty-eighth Annual ACM Symposium on
  Theory of Computing\/}, STOC '96, (pp. 212--219). New York, NY, USA:
  Association for Computing Machinery.

\bibitem[{Hagar(2007)}]{hagar2007b}
Hagar, A. (2007).
\newblock Quantum algorithms: Philosophical lessons.
\newblock {\em Minds \& Machines\/}, {\em 17\/}, 233--247.

\bibitem[{Hagar \& Cuffaro(2019)}]{hagar2019}
Hagar, A., \& Cuffaro, M. (2019).
\newblock Quantum computing.
\newblock In E.~N. Zalta (Ed.) {\em The {S}tanford Encyclopedia of
  Philosophy\/}. Metaphysics Research Lab, {S}tanford University, winter 2019
  ed.

\bibitem[{Heisenberg(1971)}]{heisenberg1971}
Heisenberg, W. (1971).
\newblock {\em Physics and Beyond\/}.
\newblock New York: Harper \& Row.

\bibitem[{Hewitt-Horsman(2009)}]{hewittHorsman2009}
Hewitt-Horsman, C. (2009).
\newblock An introduction to many worlds in quantum computation.
\newblock {\em Foundations of Physics\/}, {\em 39\/}, 869--902.

\bibitem[{Horsman et~al.(2018)Horsman, Kendon, \& Stepney}]{horsman2018}
Horsman, D., Kendon, V., \& Stepney, S. (2018).
\newblock Abstraction/representation theory and the natural science of
  computation.
\newblock In M.~E. Cuffaro, \& S.~C. Fletcher (Eds.) {\em Physical Perspectives
  on Computation, Computational Perspectives on Physics\/}, (pp. 127--152).
  Cambridge: Cambridge University Press.

\bibitem[{Howard(1989)}]{howard1989}
Howard, D. (1989).
\newblock Holism, separability, and the metaphysical implications of the {B}ell
  experiments.
\newblock In J.~T. Cushing, \& E.~McMullin (Eds.) {\em Philosophical
  Consequences of Quantum Theory\/}, (pp. 224--253). Notre Dame: University of
  Notre Dame Press.

\bibitem[{Hughes(1989)}]{hughes1989}
Hughes, R. I.~G. (1989).
\newblock {\em The Structure and Interpretation of Quantum Mechanics\/}.
\newblock Cambridge, MA.: Harvard University Press.

\bibitem[{Janas et~al.(Forthcoming)Janas, Cuffaro, \& Janssen}]{3m2020}
Janas, M., Cuffaro, M.~E., \& Janssen, M. (Forthcoming).
\newblock {\em Understanding Quantum Raffles: Quantum Mechanics on an
  Information-Theoretic Approach: Structure and Interpretation\/}.
\newblock Springer.

\bibitem[{Jozsa \& Linden(2003)}]{jozsa2003}
Jozsa, R., \& Linden, N. (2003).
\newblock On the role of entanglement in quantum-computational speed-up.
\newblock {\em Proceedings of the Royal Society of London. Series A.
  Mathematical, Physical and Engineering Sciences\/}, {\em 459\/}, 2011--2032.

\bibitem[{Kent(2005)}]{kent2005}
Kent, A. (2005).
\newblock Causal quantum theory and the collapse locality loophole.
\newblock {\em Physical Review A\/}, {\em 72\/}, 012107.

\bibitem[{Lee \& Thomas(2002)}]{lee2002}
Lee, K.~F., \& Thomas, J.~E. (2002).
\newblock Experimental simulation of two-particle quantum entanglement using
  classical fields.
\newblock {\em Physical Review Letters\/}, {\em 88\/}, 097902.

\bibitem[{Lehner(1997)}]{lehner1997}
Lehner, C. (1997).
\newblock What it feels like to be in a superposition. {A}nd why.
\newblock {\em Synthese\/}, {\em 110\/}, 191--216.

\bibitem[{Lenstra et~al.(1990)Lenstra, Lenstra~Jr., Manasse, \&
  Pollard}]{lenstra1990}
Lenstra, A.~K., Lenstra~Jr., H.~W., Manasse, M.~S., \& Pollard, J.~M. (1990).
\newblock The number field sieve.
\newblock In {\em Proceedings of the Twenty-second Annual ACM Symposium on
  Theory of Computing\/}, STOC '90, (pp. 564--572). New York, NY: Association
  for Computing Machinery.

\bibitem[{Lupacchini(2018)}]{lupacchini2018}
Lupacchini, R. (2018).
\newblock Church’s thesis, {T}uring’s limits, and {D}eutsch’s principle.
\newblock In M.~E. Cuffaro, \& S.~C. Fletcher (Eds.) {\em Physical Perspectives
  on Computation, Computational Perspectives on Physics\/}, (pp. 60--82).
  Cambridge: Cambridge University Press.

\bibitem[{Mairhofer(2021)}]{mairhofer2021}
Mairhofer, L. (2021).
\newblock {\em Atom und Individuum: Bertolt Brechts Interferenz mit der
  Quantenphysik\/}.
\newblock Berlin: De Gruyter.

\bibitem[{Maroney \& Timpson(2018)}]{maroney2018}
Maroney, O. J.~E., \& Timpson, C.~G. (2018).
\newblock How is there a physics of information? {O}n characterising physical
  evolution as information processing.
\newblock In M.~E. Cuffaro, \& S.~C. Fletcher (Eds.) {\em Physical Perspectives
  on Computation, Computational Perspectives on Physics\/}, (pp. 103--126).
  Cambridge: Cambridge University Press.

\bibitem[{Martin(1997)}]{martin1997}
Martin, J.~C. (1997).
\newblock {\em Introduction to Languages and the Theory of Computation\/}.
\newblock New York: McGraw-Hill, second ed.

\bibitem[{Maudlin(2011)}]{maudlin2011}
Maudlin, T. (2011).
\newblock {\em Quantum Non-Locality and Relativity\/}.
\newblock Cambridge, MA: Wiley-Blackwell, third ed.

\bibitem[{Mehlhorn \& Sanders(2008)}]{mehlhorn2008}
Mehlhorn, K., \& Sanders, P. (2008).
\newblock {\em Algorithms and Data Structures\/}.
\newblock Berlin: Springer.

\bibitem[{Mermin(2007)}]{mermin2007}
Mermin, D.~N. (2007).
\newblock {\em Quantum Computer Science: An Introduction\/}.
\newblock Cambridge University Press.

\bibitem[{Miranda(2021)}]{mirandaHelloMusic}
Miranda, E.~R. (2021).
\newblock Quantum computer: Hello, music!
\newblock In E.~R. Miranda (Ed.) {\em Handbook of Artificial Intelligence for
  Music: Foundations, Advanced Approaches, and Developments for Creativity\/}.
  Cham: Springer.
\newblock Forthcoming.

\bibitem[{Myrvold et~al.(2020)Myrvold, Genovese, \&
  Shimony}]{myrvoldGenoveseShimonySEP}
Myrvold, W., Genovese, M., \& Shimony, A. (2020).
\newblock Bell's theorem.
\newblock In E.~N. Zalta (Ed.) {\em The {S}tanford Encyclopedia of
  Philosophy\/}. Metaphysics Research Lab, {S}tanford University, fall 2020 ed.

\bibitem[{Nielsen(2006)}]{nielsen2006}
Nielsen, M.~A. (2006).
\newblock Cluster-state quantum computation.
\newblock {\em Reports on Mathematical Physics\/}, {\em 57\/}, 147--161.

\bibitem[{Nielsen \& Chuang(2000)}]{nielsenChuang2000}
Nielsen, M.~A., \& Chuang, I.~L. (2000).
\newblock {\em Quantum Computation and Quantum Information\/}.
\newblock Cambridge: {Cambridge University Press}.

\bibitem[{Nishimura \& Ozawa(2009)}]{nishimura2009}
Nishimura, H., \& Ozawa, M. (2009).
\newblock Perfect computational equivalence between quantum turing machines and
  finitely generated uniform quantum circuit families.
\newblock {\em Quantum Information Processing\/}, {\em 8\/}, 13--24.

\bibitem[{Pearl(2009)}]{pearl2009}
Pearl, J. (2009).
\newblock {\em Causality: Models, Reasoning, and Inference\/}.
\newblock Cambridge: Cambridge University Press, 2nd ed.

\bibitem[{Pitowsky(1989)}]{pitowsky1989}
Pitowsky, I. (1989).
\newblock {\em Quantum Probability --- Quantum Logic\/}.
\newblock Hemsbach: Springer.

\bibitem[{Pitowsky(1994)}]{pitowsky1994}
Pitowsky, I. (1994).
\newblock {G}eorge {B}oole's `conditions of possible experience' and the
  quantum puzzle.
\newblock {\em British Journal for the Philosophy of Science\/}, {\em 45\/},
  99--125.

\bibitem[{Pitowsky(2002)}]{pitowsky2002}
Pitowsky, I. (2002).
\newblock Quantum speed-up of computations.
\newblock {\em Philosophy of Science\/}, {\em 69\/}, S168--S177.

\bibitem[{Raussendorf \& Briegel(2002)}]{raussendorf2002}
Raussendorf, R., \& Briegel, H.~J. (2002).
\newblock Computational model underlying the one-way quantum computer.
\newblock {\em Quantum Information and Computation\/}, {\em 2\/}, 443--486.

\bibitem[{Raussendorf et~al.(2003)Raussendorf, Browne, \&
  Briegel}]{raussendorf2003}
Raussendorf, R., Browne, D.~E., \& Briegel, H.~J. (2003).
\newblock Measurement-based quantum computation on cluster states.
\newblock {\em Physical Review A\/}, {\em 68\/}, 022312.

\bibitem[{Rivest et~al.(1978)Rivest, Shamir, \& Adleman}]{rsa1978}
Rivest, R.~L., Shamir, A., \& Adleman, L. (1978).
\newblock A method for obtaining digital signatures and public-key
  cryptosystems.
\newblock {\em Communications of the ACM\/}, {\em 21\/}, 120--126.

\bibitem[{Rosset et~al.(2013)Rosset, Branciard, Gisin, \& Liang}]{rosset2013}
Rosset, D., Branciard, C., Gisin, N., \& Liang, Y.-C. (2013).
\newblock Entangled states cannot be classically simulated in generalized
  {B}ell experiments with quantum inputs.
\newblock {\em New Journal of Physics\/}, {\em 15\/}, 053025.

\bibitem[{Saunders(1995)}]{saunders1995}
Saunders, S. (1995).
\newblock Time, quantum mechanics, and decoherence.
\newblock {\em Synthese\/}, {\em 102\/}, 235--266.

\bibitem[{Schr\"odinger(1935{\natexlab{a}})}]{schrodinger1935a}
Schr\"odinger, E. (1935{\natexlab{a}}).
\newblock Die gegenw\"artige situation in der quantenmechanik.
\newblock {\em Naturwissenschaften\/}, {\em 23\/}, 807--812; 823--828;
  844--849.
\newblock Translated in: John D. Trimmer (1980) in \emph{Proceedings of the
  American Philosophical Society 124, 323--338}.

\bibitem[{Schr\"odinger(1935{\natexlab{b}})}]{schrodinger1935}
Schr\"odinger, E. (1935{\natexlab{b}}).
\newblock Discussion of probability relations between separated systems.
\newblock {\em Mathematical Proceedings of the Cambridge Philosophical
  Society\/}, {\em 31\/}, 555--563.

\bibitem[{Shetterly(2016)}]{shetterly2016}
Shetterly, M.~L. (2016).
\newblock {\em Hidden Figures\/}.
\newblock New York: HarperCollins.

\bibitem[{Shin et~al.(2014)Shin, Smith, Smolin, \& Vazirani}]{shin2014}
Shin, S.~W., Smith, G., Smolin, J.~A., \& Vazirani, U. (2014).
\newblock How "quantum" is the {D-Wave} machine?
\newblock {arXiv:1401.7087v2}.

\bibitem[{Shor(1994)}]{shor1994}
Shor, P.~W. (1994).
\newblock Algorithms for quantum computation: Discrete logarithms and
  factoring.
\newblock {\em Foundations of Computer Science, 1994 Proceedings., 35th Annual
  Symposium on\/}, (pp. 124--134).

\bibitem[{Sipser(1992)}]{sipser1992}
Sipser, M. (1992).
\newblock The history and status of the {P} versus {NP} question.
\newblock In {\em Proceedings of the Twenty-fourth Annual ACM Symposium on
  Theory of Computing\/}, STOC '92, (pp. 603--618). New York, NY: Association
  for Computing Machinery.

\bibitem[{Tessier(2004)}]{tessier2004}
Tessier, T.~E. (2004).
\newblock {\em Complementarity and Entanglement in Quantum Information
  Theory\/}.
\newblock Ph.D. thesis, The University of New Mexico, Albuquerque, New Mexico.

\bibitem[{Tessier et~al.(2005)Tessier, Caves, Deutsch, \& Eastin}]{tessier2005}
Tessier, T.~E., Caves, C.~M., Deutsch, I.~H., \& Eastin, B. (2005).
\newblock Optimal classical-communication-assisted local model of $n$-qubit
  {Greenberger-Horne-Zeilinger} correlations.
\newblock {\em Physical Review A\/}, {\em 72\/}, 032305.

\bibitem[{Timpson(2013)}]{timpson2013}
Timpson, C.~G. (2013).
\newblock {\em Quantum Information Theory \& the Foundations of Quantum
  Mechanics\/}.
\newblock Oxford: Oxford University Press.

\bibitem[{Toner \& Bacon(2003)}]{toner2003}
Toner, B.~F., \& Bacon, D. (2003).
\newblock Communication cost of simulating {B}ell correlations.
\newblock {\em Physical Review Letters\/}, {\em 91\/}, 187904.

\bibitem[{Turing(1936-7)}]{turing1937}
Turing, A.~M. (1936-7).
\newblock On computable numbers, with an application to the
  {E}ntscheidungsproblem.
\newblock {\em Proceedings of the London Mathematical Society. Second
  Series\/}, {\em s2-42\/}, 230--265.

\bibitem[{Vaidman(1998)}]{vaidman1998}
Vaidman, L. (1998).
\newblock On schizophrenic experiences of the neutron or why we should believe
  in the many‐worlds interpretation of quantum theory.
\newblock {\em International Studies in the Philosophy of Science\/}, {\em
  12\/}, 245--261.

\bibitem[{Vaidman(2012)}]{vaidman2012}
Vaidman, L. (2012).
\newblock Probability in the many-worlds interpretation of quantum mechanics.
\newblock In Y.~Ben-Menahem, \& M.~Hemmo (Eds.) {\em Probability in Physics\/},
  (pp. 299--311). Berlin: Springer.

\bibitem[{Vaidman(2018 [2002])}]{vaidman2018}
Vaidman, L. (2018 [2002]).
\newblock Many-worlds interpretation of quantum mechanics.
\newblock In E.~N. Zalta (Ed.) {\em The {S}tanford Encyclopedia of
  Philosophy\/}. Metaphysics Research Lab, {S}tanford University, fall 2018 ed.
\newblock First published: 2002.

\bibitem[{van Emde~Boas(1990)}]{boas1990}
van Emde~Boas, P. (1990).
\newblock Machine models and simulations.
\newblock In J.~van Leeuwen (Ed.) {\em Handbook of Theoretical Computer
  Science, Volume {A}: Algorithms and Complexity\/}, (pp. 1--66). Cambridge,
  MA: MIT Press/Elsevier.

\bibitem[{Vincent(1951)}]{vincent1951}
Vincent, J. (1951).
\newblock {\em The Diatonic Modes in Modern Music\/}.
\newblock New York: Mills Music.
\newblock Page references to second edition (1974) published by Curlew Music
  Publishers, Hollywood, CA.

\bibitem[{Wallace(2003)}]{wallace2003}
Wallace, D. (2003).
\newblock Everett and structure.
\newblock {\em Studies in History and Philosophy of Modern Physics\/}, {\em
  34\/}, 87--105.

\bibitem[{Wallace(2007)}]{wallace2007}
Wallace, D. (2007).
\newblock Quantum probability from subjective likelihood: Improving on
  {D}eutsch's proof of the probability rule.
\newblock {\em Studies in History and Philosophy of Modern Physics\/}, {\em
  38\/}, 311--332.

\bibitem[{Wallace(2010)}]{wallace2010}
Wallace, D. (2010).
\newblock Decoherence and ontology.
\newblock In S.~Saunders, J.~Barrett, A.~Kent, \& D.~Wallace (Eds.) {\em Many
  Worlds? Everett, Quantum Theory, and Reality\/}, (pp. 53--72). Oxford: Oxford
  University Press.

\bibitem[{Wallace(2012)}]{wallace2012}
Wallace, D. (2012).
\newblock {\em The Emergent Multiverse\/}.
\newblock Oxford: Oxford University Press.

\bibitem[{Wallace(2019)}]{wallace2019}
Wallace, D. (2019).
\newblock On the plurality of quantum theories: Quantum theory as a framework,
  and its implications for the quantum measurement problem.
\newblock In S.~French, \& J.~Saatsi (Eds.) {\em Realism and the Quantum\/},
  (pp. 78--102). Oxford: Oxford University Press.

\bibitem[{{Wikipedia contributors}(2020)}]{wikiMips}
{Wikipedia contributors} (2020).
\newblock Instructions per second.
\newblock In {\em Wikipedia, The Free Encyclopedia\/}.
\newblock Posted on 15 September, 2020, 11:15.

\bibitem[{Zurek({2003 [1991]})}]{zurek2003}
Zurek, W.~H. ({2003 [1991]}).
\newblock Decoherence and the transition from quantum to classical --
  revisited.
\newblock {arXiv:quant-ph/0306072v1}.

\end{thebibliography}

\end{document}